\documentclass{aa}
\usepackage {graphicx}
\usepackage {amssymb}
\usepackage {mathrsfs}
\usepackage {txfonts}
\def\Teff  {T$_{\mbox {\scriptsize eff}}$}

\def\vt  {$v_t$}
\def\kms   {$ {\rm km \: s^{-1}  } $}
\def\12c13 {$ {\rm ^{12}C/^{13}C } $}
\def\12c   {$ {\rm ^{12}C  } $}
\def\13c   {$ {\rm ^{13}C  } $}

\begin {document}


\title {First stars IX - Mixing in extremely metal-poor giants. 
Variation of the {$ {\rm ^{12}C/^{13}C } $}, [Na/Mg] and [Al/Mg]
~ratios.
\thanks{Based on observations obtained with the ESO Very Large
Telescope at Paranal Observatory, Chile (Large Programme ``First
Stars'', ID 165.N-0276(A); P.I.: R. Cayrel).}
}

\author {
M. Spite\inst{1}\and 
R. Cayrel\inst{1}\and
V. Hill\inst{1}\and
F. Spite\inst{1}\and
P. Fran\c cois\inst{1}\and
B. Plez\inst{2}\and
P. Bonifacio\inst{1}\and
P. Molaro\inst{1,4}\and
E. Depagne\inst{3}\and 
J. Andersen\inst{7,8}\and
B. Barbuy\inst{5}\and
T.C. Beers\inst{6}\and
B. Nordstr\"{o}m\inst{7,9}\and
F. Primas\inst{10}
}

\offprints {monique.spite@obspm.fr}

\institute {
   GEPI, Observatoire de Paris-Meudon, F-92125 Meudon Cedex, France,
\and 
   GRAAL, Universit\'e de Montpellier II, F-34095 Montpellier Cedex 
05, France,
\and
   European Southern Observatory (ESO), Casilla 19001, Santiago 19,
Chile
\and
    Osservatorio Astronomico di Trieste, INAF,
    Via G.B. Tiepolo 11, I-34131 Trieste, Italy,
\and
    IAG, Universidade de Sao Paulo, Depto. de Astronomia, 
    Rua do Matao 1226, Sao Paulo 05508-900, Brazil, 
\and
Department of Physics \& Astronomy, CSCE: Center for the Study of
Cosmic
Evolution, and JINA: Joint Institute for Nuclear
Astrophysics, Michigan State University, East Lansing, MI 48824, USA
\and
  The Niels Bohr Institute, Astronomy; Juliane Maries Vej 30, 
  DK-2100 Copenhagen, Denmark,
\and
  Nordic Optical Telescope Scientific Association, Apartado 474, 
  ES-38 700 Santa Cruz de La Palma, Spain,
\and
  Lund Observatory, Box 43, SE-221 00 Lund, Sweden,
\and
  European Southern Observatory, Karl Schwarzschild-Str. 2, 
  D-85749 Garching bei M\"unchen, Germany
}

\date {Received XXX; accepted XXX}
\titlerunning {${\rm ^{12}C/^{13}C}$ ~in extremely metal-poor giants}
\authorrunning {M. Spite et al.}

\abstract
{Extremely metal-poor (EMP) stars preserve a fossil record of the
composition of the ISM when the Galaxy formed.  It is crucial,
however, to verify whether internal mixing has modified their surface
composition, especially in the giants where most elements can be
studied.
}
{We aim to understand the CNO abundance variations found in some, but
not all EMP field giants analysed earlier.  Mixing beyond the first
dredge-up of standard models is required, and its origin needs
clarification.
}
{The ${\rm ^{12}C/^{13}C}$ ratio is the most robust diagnostic of deep
mixing, because it is insensitive to the adopted stellar parameters
and should be uniformly high in near-primordial gas.  We have measured
${\rm ^{12}C ~and~ ^{13}C}$ abundances in 35 EMP giants (including 22
with ${\rm[Fe/H] < -3.0}$) from high-quality VLT/UVES spectra analysed
with LTE model atmospheres.  Correlations with other abundance data
are used to study the depth of mixing.
}
{The ${\rm ^{12}C/^{13}C}$ ratio is found to correlate with [C/Fe] (and
Li/H), and clearly anti-correlate with [N/Fe], as expected if the
surface abundances are modified by CNO processed material from the
interior.  Evidence for such deep mixing is observed in giants above
${\rm log~L/L_{\odot} = 2.6}$, brighter than in less metal-poor stars,
but matching the bump in the luminosity function in both cases.  Three
of the mixed stars are also Na- and Al-rich, another signature of deep
mixing, but signatures of the ON cycle are not clearly seen in these
stars.
}
{Extra mixing processes clearly occur in luminous RGB stars.  They
cannot be explained by standard convection, nor in a simple way by
rotating models.  The Na- and Al-rich giants could be AGB stars
themselves, but an inhomogeneous early ISM or pollution from a binary
companion remain possible alternatives.
}
\maketitle
%
\begin {figure*}[ht]
\begin {center}
\resizebox  {12.0cm}{6.0cm} 
{\includegraphics {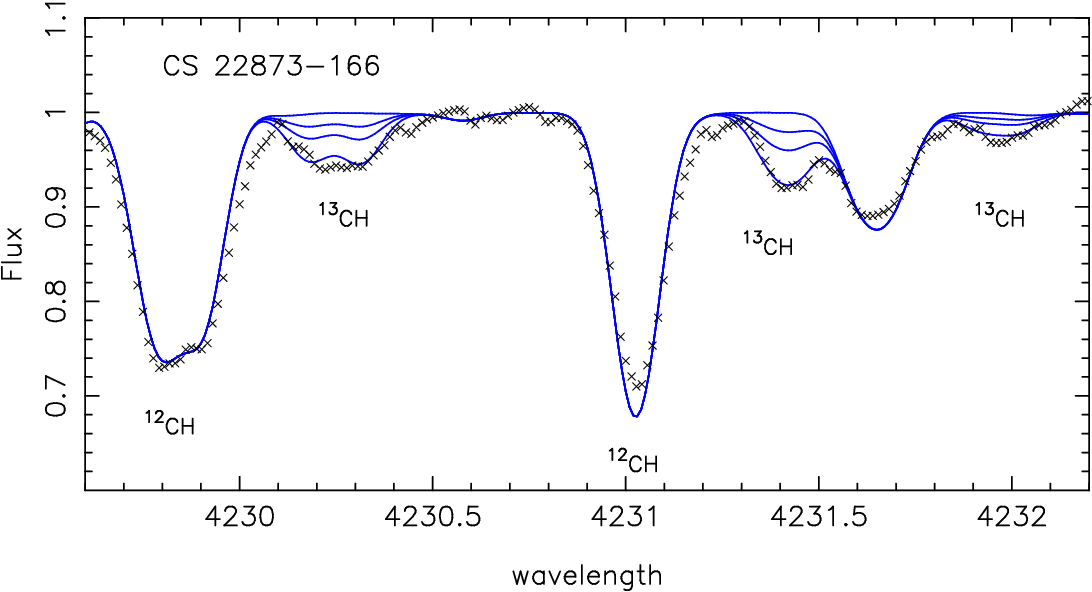} }
\resizebox  {12.0cm}{6.0cm} 
{\includegraphics {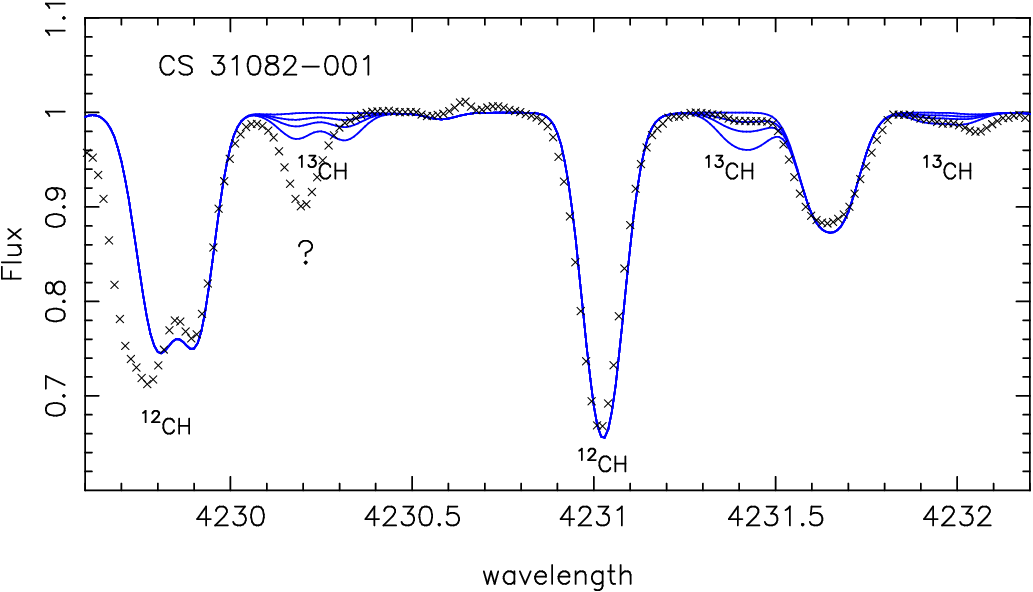} }
\caption {Reduced UVES spectra (crosses) in the region of the ${\rm
^{13}C}$ features.  Synthetic spectra are shown (full lines), computed
without ${\rm ^{13}C}$ as well as for $\log {\rm ^{12}C/^{13}C }$ =
1.3 1.0 0.7 for CS~22873- 166, $\log {\rm ^{12}C/^{13}C}$ = 1.6 1.3
1.0 for CS~31082-001.  The best fits are for ${\rm log^{12}C/^{13}C} $
= 0.7 ({${\rm ^{12}C/^{13}C }$} = 5) in CS~22873-166 and {${\rm
log^{12}C/^{13}C}$} = 1.6 ({${\rm ^{12}C/^{13}C}$} = 40) in
CS~31082-001.  The ${\rm ^{13}C}$ line at 4230.3\AA~ in CS~31082-001
is
severely blended by an unidentified line, probably due to an
$r$-process element, as these elements are strongly enhanced in this
star (Hill et al.  \cite{HPC02}).
}
\label {spectra}
\end {center}
\end {figure*}

\section {Introduction} 

The surface composition of a cool star is a good diagnostic of the
chemical composition of the gas from which it formed, if mixing with
material processed inside the star itself has not occurred.  We are
conducting a comprehensive spectroscopic study of extremely metal-poor
(EMP) stars to obtain precise and detailed information on the chemical
composition of the ISM in the early Galaxy and the yields of the first
generation(s) of supernovae.  Using the ESO VLT and UVES spectrograph,
we have obtained abundance results of unprecedented quality for a
large sample of faint halo stars.  The next step is to investigate to
what extent these abundances represent the composition of the early
ISM.

The most complete range of elements can be studied in the cool,
low-gravity atmospheres of red giant stars.  Accordingly, Cayrel et
al.  (\cite{CDS04}; ``First Stars V'') determined abundances for the
elements from C to Zn in a sample of 35 EMP giants.  30 of the stars
were found to have ${\rm -4.1 <[Fe/H]< -2.7}$, while 22 stars have
${\rm[Fe/H] \leqslant -3.0}$.  Very tight abundance relations were
found for nearly all elements, with C and N as notable exceptions.  As
C, N, and O are the first elements synthesised after the Big Bang,
their abundances in the early galactic gas are of particular interest
and merit further study.

In a subsequent paper (Spite et al., \cite{SCP05}; ``First Stars
VI''), we found the C and N abundances in our sample to be
anti-correlated.  Two groups of stars were clearly separated in a plot
of [N/Fe] vs.  [C/Fe]: A first group, named ``unmixed'' stars had
${\rm [N/Fe] < 0.5}$, ${\rm [C/Fe] > 0.0}$, and measurable Li
abundances in the range ${\rm 0.2 < A(Li) < 1.2}$, while a second
group of ``mixed'' stars had ${\rm [N/Fe] > 0.5}$, ${\rm [C/Fe] <
0.0}$, and Li below our detection threshold.  However, [(C+N)/Fe] was
practically identical in the two subsamples.  We also found the
``unmixed'' stars to lie mostly on the lower part of the Red Giant
Branch (RGB), while the ``mixed'' stars seemed to lie on the upper RGB
or even the Horizontal Branch (HB) or asymptotic giant branch (AGB).
Let us remark here that the adjectives ``mixed" and ``unmixed", as
defined in our paper ``First Stars VI", may appear a bit misleading.  
All of the stars studied in these papers are giants, and thus they all have
undergone the first dredge-up.  This can be clearly seen from the
abundance of lithium, which is close to A(Li) = 2.3 in EMP dwarfs (Bonifacio
et al., \cite{BMS06}, ``First Stars VII") and is only ${\rm A(Li)} < 1.2$ 
in our sample of giants (see ``First Stars VI").  Lithium is a very
fragile element, and mixing with layers at temperatures higher than
only $2.5~10^6$K is able to destroy it.  As a first approximation, we
can suppose that the first dredge-up does not affect the abundances of
the elements heavier than Li in low-mass stars ($\mathscr{M/M}_{\odot}
\leq 0.9$).  The effect on [C/Fe] and [N/Fe], following, e.g.,
Gratton et al. (\cite{GSC00}) or Denissenkov \& Weiss (\cite{DW04})
is negligeable.  In ``First stars VI", and in this paper we use the term ``mixed
stars" only for stars that have undergone mixing with deep layers of the star,
thereby affecting the observed CNO abundances, and ``unmixed" stars for those
which such deep mixing is not expected to have occured.

The simplest explanation of the observed C and N abundances is that
mixing has occurred between the atmosphere of the upper RGB stars and
the H-burning layer where C is converted into N by the CNO cycle
(e.g. Charbonnel \cite{Ch95}).  Evidence for mixing in upper RGB
stars has already been observed in globular cluster giants (Kraft
\cite{Kr94},  Bellman et al. \cite{BBS01}, Shetrone \cite{Sh03}, Grundahl et al.
\cite{GBN02}), open clusters (Gilroy
\cite{Gi89}, Jacobson et al. \cite{JPF05}), and field giants
(Shetrone et al., \cite{SSP93}, Charbonnel et al.  \cite{CBW98},
Gratton et al. \cite{GSC00}).  However, these stars were only
moderately metal-poor (${\rm -2 <[Fe/H]< -0.5}$), substantially more
metal-rich than our present sample.

Convection, the only mechanism of internal mixing in ``standard
models", cannot by itself account for these observations.  Different
models have been proposed to explain the observed extra mixing (e.g.
Denissenkov \& VandenBerg \cite{DV03}; Chanam\'e et al.  
\cite{CPT05}, Palacios et al. \cite{PCT06})
and predict that mixing should be a function of metallicity: at any
given depth in a stellar interior, a metal-poor star is hotter than
a metal-rich star of the same mass; thus, the CN and ON processing
shells are closer to the surface, facilitating mixing (Charbonnel
\cite{Ch94}; Charbonnel et al., \cite{CBW98}).  However, Gratton et
al.  (\cite{GSC00}) note that, in their sample of moderately
metal-poor stars, mixing apparently reaches a maximum for stars with
${\rm [Fe/H] \approx -1.5}$.

The primary aim of the present paper is to better constrain the models
of internal mixing in giant stars, especially in EMP giants.
Measurements of the {${\rm ^{12}C/^{13}C}$} ~ratio are a particularly
powerful tool, because {${\rm ^{12}C/^{13}C}$} is largely unaffected
by uncertainties in the adopted stellar parameters, and appears to be
high ($>70$) in nearly primordial gas (e.g. Levshakov et al.
\cite{LCM05}).  As a result, any significant variation of {${\rm
^{12}C/^{13}C}$} should be due to internal mixing processes in the
stars.

\section {Observations and reduction}

The observations were performed with the ESO VLT and its
high-resolution spectrograph UVES (Dekker et al.  \cite {DDK00}).
Briefly, the spectra have a resolution of $R$= 47,000 at $\lambda$ =
400 nm and typical S/N ratios per pixel of $\sim$130, with an average
of 5 pixels per resolution element; they were reduced using the UVES
context within MIDAS (Ballester et al.  \cite {BMB00}).  Our spectra
and their reduction are described in detail by Cayrel et al.
(\cite{CDS04}; ``First Stars V"); sample spectra of the region of
interest in this paper are shown in Fig.  \ref{spectra}.

\section {Analysis}  \label{analysis}

We have carried out a classical LTE analysis using OSMARCS models
(see, e.g., Gustafsson et al.  \cite{GBE75}, \cite{GEE03}).  \Teff\
was derived from photometry, using the calibration of Alonso et al.
\cite{AAM99}, and log g from the ionisation equilibrium of Fe and Ti.
Microturbulent velocities were fixed by requiring no trend of 
[Fe~{\footnotesize I}/H]
with equivalent width.  Details of the analysis are given in ``First
Stars V".

\begin {table*}[t]
\caption {Adopted model atmosphere parameters (\Teff, log $g$, \vt,
[Fe/H]), C and N abundances, and ${\rm ^{12}C/^{13}C}$ ratios for the
programme stars.  [C/H], as given in this table, is the sum ${\rm
[(^{12}C+^{13}C)/H]}$.  An ``m'' in the ``Mix'' column denotes the
``mixed'' stars.  The two ``C -rich" stars are very peculiar and are
omitted from the figures and the discussion.
}
\label {tabund}
\begin {center}
\begin{tabular}{llc@{ }c@{ }c@{ } c@{ }c@{ }c@{ }c@{ }c@{ }r@{ }r@{
}c@{ }c@{ }c@{ }c}
\hline 
\hline
&Star          &   $T_{eff}$& ~log g~&~\vt~&~[Fe/H]~ & log L/L$_\odot$ &[C/H] & [N/H]  & [(C+N)/H]  & [C/N]
&${\rm^{12}C/^{13}C}$ & mix. & Rem. \\
\hline
~1 & HD~2796        & 4950 & 1.5 & 2.1 & -2.47 & 2.60 &$ -2.87\pm
0.06$&$ -1.62\pm 0.08$& -2.23  & -1.25   & $4 \pm   2 $  & ~m~ & \\
~2 & HD~122563      & 4600 & 1.1 & 2.0 & -2.82 & 2.87 &$ -3.21\pm
0.05$&$ -2.12\pm 0.15$& -2.70  & -1.09   & $5 \pm   2 $  & ~m~ & \\
~3 & HD~186478      & 4700 & 1.3 & 2.0 & -2.59 & 2.71 &$ -2.81\pm
0.07$&$ -1.97\pm 0.12$& -2.47  & -0.84   & $5 \pm   2 $  & ~m~ & \\
~4 & BD~+17:3248    & 5250 & 1.4 & 1.5 & -2.07 & 2.80 &$ -2.40\pm
0.05$&$ -1.42\pm 0.10$& -1.97  & -0.98   & $10\pm   5 $  & ~m~ & \\
~5 & BD~--18:5550   & 4750 & 1.4 & 1.8 & -3.06 & 2.63 &$ -3.08\pm
0.04$&$ -3.42\pm 0.10$& -3.13  & +0.34   & $>40$         &     & \\
~6 & CD~--38:245    & 4800 & 1.5 & 2.2 & -4.19 & 2.55 &$<-4.52
$&$ -3.12\pm 0.20$&$<-3.75$&$<-1.40$ &   -           & ~m~ & \\
~7 & BS~16467--062  & 5200 & 2.5 & 1.6 & -3.77 & 1.69 &$ -3.52\pm
0.12$&$ <-3.32       $&$<-3.47$&$>-0.20$ &   -           &     & \\
~8 & BS~16477--003  & 4900 & 1.7 & 1.8 & -3.36 & 2.38 &$ -3.07\pm
0.08$&$ <-3.62       $&$<-3.14$&$>+0.55$ & $>30$         &     & \\
~9 & BS~17569--049  & 4700 & 1.2 & 1.9 & -2.88 & 2.81 &$ -2.93\pm
0.05$&$ -2.02\pm 0.12$& -2.54  & -0.91   & $6   \pm   2$ & ~m~ & \\
10 & CS~22169--035  & 4700 & 1.2 & 2.2 & -3.04 & 2.81 &$ -3.20\pm
0.05$&$ -2.02\pm 0.13$& -2.62  & -1.18   & $6.5 \pm   2$ & ~m~ & \\
11 & CS~22172--002  & 4800 & 1.3 & 2.2 & -3.86 & 2.75 &$ -3.86\pm
0.10$&$ -3.62\pm 0.20$& -3.80  & -0.24   & $>10$         &     & \\
12 & CS~22186--025  & 4900 & 1.5 & 2.0 & -3.00 & 2.58 &$ -3.54\pm
0.10$&$ -2.02\pm 0.08$& -2.67  & -1.52   &   -           & ~m~ & \\
13 & CS~22189--009  & 4900 & 1.7 & 1.9 & -3.49 & 2.38 &$ -3.15\pm
0.08$&$ -3.22\pm 0.12$& -3.16  & +0.07   & 15 (+8 -5)    &     & \\
14 & CS~22873--055  & 4550 & 0.7 & 2.2 & -2.99 & 3.25 &$ -3.62\pm
0.06$&$ -1.92\pm 0.20$& -2.58  & -1.70   & $4  \pm   2$  & ~m~ &
Na-rich\\
15 & CS~22873--166  & 4550 & 0.9 & 2.1 & -2.97 & 3.05 &$ -3.10\pm
0.08$&$ -1.92\pm 0.20$& -2.52  & -1.18   & $5  \pm   2$  & ~m~ & \\
16 & CS~22878--101  & 4800 & 1.3 & 2.0 & -3.25 & 2.75 &$ -3.46\pm
0.10$&$ -1.92\pm 0.10$& -2.57  & -1.54   & $5  \pm   2$  & ~m~ & \\
17 & CS~22885--096  & 5050 & 2.6 & 1.8 & -3.78 & 1.53 &$ -3.52\pm
0.06$&$ -3.52\pm 0.13$& -3.52  &  0.00   &   -           &     & \\
18 & CS~22891--209  & 4700 & 1.0 & 2.1 & -3.29 & 3.01 &$ -3.86\pm
0.05$&$ -2.17\pm 0.10$& -2.83  & -1.69   & $5  \pm   2$  & ~m~ &
Na-rich\\
19 & CS~22892--052  & 4850 & 1.6 & 1.9 & -3.03 & 2.46 &$ -2.11\pm
0.06$&$ -2.52\pm 0.13$& -2.17  & +0.41   & 16 (+8 -5)    &     &
C-rich\\
20 & CS~22896--154  & 5250 & 2.7 & 1.2 & -2.69 & 1.50 &$ -2.46\pm
0.05$&$ -2.92\pm 0.12$& -2.52  & +0.46   & $>40$         &     & \\
21 & CS~22897--008  & 4900 & 1.7 & 2.0 & -3.41 & 2.38 &$ -2.83\pm
0.05$&$ -3.17\pm 0.15$& -2.88  & +0.34   &~20 (+8 -5)    &     & \\
22 & CS~22948--066  & 5100 & 1.8 & 2.0 & -3.14 & 2.35 &$ -3.06\pm
0.10$&$ -1.92\pm 0.10$& -2.51  & -1.14   & -             & ~m~ & \\
23 & CS~22949--037  & 4900 & 1.5 & 1.8 & -3.97 & 2.58 &$ -2.72\pm
0.10$&$ -1.72\pm 0.30$& -2.27  & -1.00   & $4  \pm   2$  & ~m~ &
C-rich\\
24 & CS~22952--015  & 4800 & 1.3 & 2.1 & -3.43 & 2.75 &$ -4.02\pm
0.08$&$ -2.12\pm 0.10$& -2.80  & -1.90   &   -           & ~m~ &
Na-rich\\
25 & CS~22953--003  & 5100 & 2.3 & 1.7 & -2.84 & 1.85 &$ -2.52\pm
0.03$&$ -2.72\pm 0.10$& -2.55  & +0.20   & 20  (+8 -5)   &     & \\
26 & CS~22956--050  & 4900 & 1.7 & 1.8 & -3.33 & 2.38 &$ -3.06\pm
0.05$&$ -3.02\pm 0.10$& -3.05  & -0.04   &   -           &     & \\
27 & CS~22966--057  & 5300 & 2.2 & 1.4 & -2.62 & 2.02 &$ -2.56\pm
0.05$&$ -2.52\pm 0.12$& -2.55  & -0.04   &   -           &     & \\
28 & CS~22968--014  & 4850 & 1.7 & 1.9 & -3.56 & 2.36 &$ -3.30\pm
0.06$&$ -3.32\pm 0.10$& -3.30  & +0.02   & 30  (+8 -5)   &     & \\
29 & CS~29491--053  & 4700 & 1.3 & 2.0 & -3.04 & 2.71 &$ -3.25\pm
0.05$&$ -2.22\pm 0.15$& -2.78  & -1.03   & $7 \pm   2$   & ~m~ & \\
30 & CS~29495--041  & 4800 & 1.5 & 1.8 & -2.82 & 2.55 &$ -2.86\pm
0.06$&$ -2.42\pm 0.10$& -2.73  & -0.44   & 14 (+8 -5)    &     & \\
31 & CS~29502--042  & 5100 & 2.5 & 1.5 & -3.19 & 1.65 &$ -3.03\pm
0.04$&$ -3.62\pm 0.20$& -3.10  & +0.59   & $>30$         &     & \\
32 & CS~29516--024  & 4650 & 1.2 & 1.7 & -3.06 & 2.79 &$ -3.10\pm
0.05$&$ -3.82\pm 0.20$& -3.18  & +0.72   & 20 (+8 -5)    &     & \\
33 & CS~29518--051  & 5200 & 2.6 & 1.4 & -2.69 & 1.59 &$ -2.77\pm
0.05$&$ -1.87\pm 0.15$& -2.39  & -0.90   & $8 \pm   2$   & ~m~ & \\
34 & CS~30325--094  & 4950 & 2.0 & 1.5 & -3.30 & 2.10 &$ -3.28\pm
0.05$&$ -3.12\pm 0.18$& -3.24  & -0.16   & 20 (+8 -5)    &     & \\
35 & CS~31082--001  & 4825 & 1.5 & 1.8 & -2.91 & 2.56 &$ -2.68\pm
0.05$&$ -3.42\pm 0.10$& -2.76  & +0.74   & $>30$         &     & \\
\hline 
\end {tabular}
\end {center}
\end {table*} 

The C abundance and ${\rm ^{12}C/^{13}C}$ ratio were determined by
spectrum synthesis of the $A^{2}\Delta - X^{2}\Pi$ band of CH (the G
band), using the current version of the Turbospectrum code (Alvarez \&
Plez \cite{AP98}).  The CH line positions were computed with the
program LIFBASE (Luque \& Crosley \cite{LC99}), and $gf$ values,
excitation energies, and isotopic shifts were taken from the line list
of J{\o}rgensen et al.  (\cite{JLI96}).  Several features of the band
were used, but the stronger lines of ${\rm ^{13}C}$ are at 423.030,
423.145, and 423.655 nm, respectively.  Sample spectrum fits are shown
in Fig.  \ref{spectra}.

We note that the ${\rm ^{13}C}$ line at 423.030 nm is sometimes
blended by an unidentified line, probably due to an $r$-process
element.  In the extreme $r$- process-element rich giant CS~31082-001
(Hill et al.  \cite{HPC02}, ``First Stars I''), this line is quite
strong (Fig.  \ref{spectra}), and it may contribute significantly to
the ${\rm ^{13}C}$ feature in other $r$-process enhanced EMP giants.
Alternative regions of the spectra of r-process-rich stars, such as
the near-IR, might be usefully explored for measurement of this
species.

Table \ref{tabund} lists the adopted parameters for our program stars
as well as the derived N abundance (from ``First Stars VI''), C
abundance and ${\rm ^{12}C/^{13}C}$ ratios.  Note that [C/H] as given
in Table~\ref{tabund} refers to the total C abundance 
(${\rm ^{12}C+^{13}C}$).  It may be slightly different from the values given in
``First Stars VI'', where [C/H] was derived from the $ {\rm ^{12}CH}$
lines, assuming a solar ${\rm ^{12}C/^{13}C}$ ratio, but the maximum
difference is less than 2\%.

Deriving ${\rm ^{13}C}$ abundances in our very weak-lined EMP stars is
a challenge, as the parent {${\rm ^{12}CH} $} ~lines are already quite
weak.  [C/H] is relatively high in the unmixed stars, but
${\rm^{13}C/^{12}C}$ is then low and the ${\rm ^{13}C}$ lines weak.
And while ${\rm^{13}C/^{12}C}$ is higher in the mixed stars,
[$^{12}$C/H] itself is low because some of the C has been transformed
into N. As a result, the measurement errors given in Table
\ref{tabund} are relatively large; for eight stars we could not
measure the ${\rm^{12}C/^{13}C}$ ratio at all.  In three of them
(CD~--38:245, BS~16467-- 062, and CS~22952--015), [C/H] is so low that
no parent ${\rm^{12}CH}$ lines are visible in the region of the
${\rm^{13}CH}$ lines.  For the five other stars, the measured upper
limit of ${\rm^{12}C/^{13}C}$ is not very useful (${\rm^{12}C/^{13}C >
5}$).

Fortunately, uncertainties in the physical parameters of the stars do
not significantly affect the resulting ${\rm^{12}C/^{13}C}$ ratio:
differences in \Teff, log g, and even microturbulent velocity affect
the ${\rm^{12}CH}$ and ${\rm^{13}CH}$ molecules nearly identically,
especially when the lines are of similar strength and/or not
saturated, as in our EMP giants (see, e.g., Sneden et al.
\cite{SPV86}).  3D effects should also be unimportant for the isotope
ratio.  Therefore, the errors of ${\rm ^{12}C/^{13}C}$ as given in
Table \ref{tabund} should reflect the total uncertainty in the isotope
ratio.

Two of our EMP giants are ``C-rich'' (Table \ref{tabund}, last
column); see Depagne et al.  (\cite{DHS02}) and Sneden et al.
(\cite{SCI00}, \cite{SCL03}).  These very peculiar stars are omitted
in the following discussion.

In ``First Stars I'', we found ${\rm ^{12}C/^{13}C > 20}$ for
CS~31082-001, but the additional spectra obtained by Plez et al.
(\cite{PHC04}) permit us to tighten this upper limit: ${\rm
^{12}C/^{13}C > 30}$.  Unfortunately, in CS~31082-001 the accuracy of
the determination is affected by an unidentified line blending with
the strongest ${\rm ^{13}C}$ line (see Fig.  \ref{spectra} and the
text above).

\begin {figure}[h]
\begin {center}
\resizebox  {8.0cm}{4.5cm} 
{\includegraphics {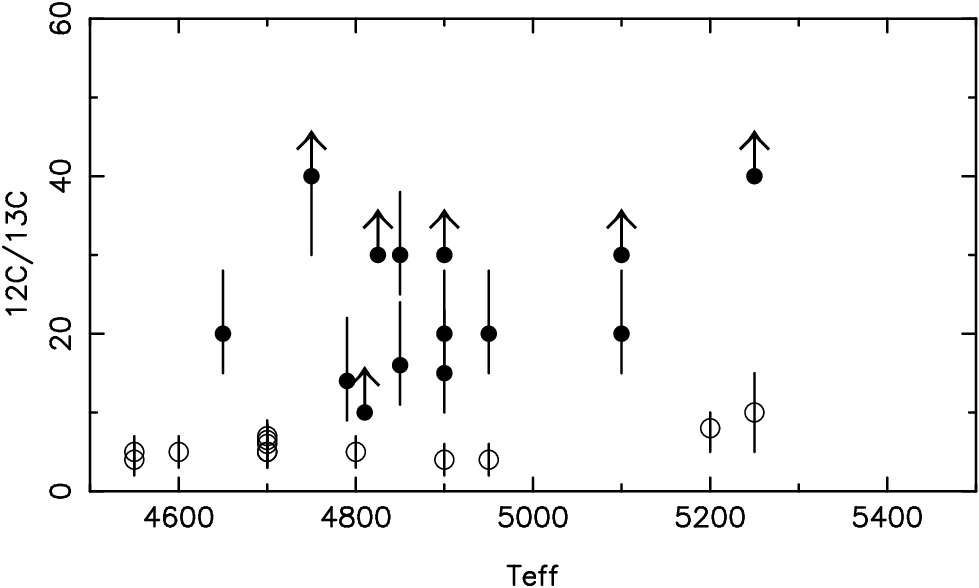} }
\resizebox  {8.0cm}{4.5cm}
{\includegraphics {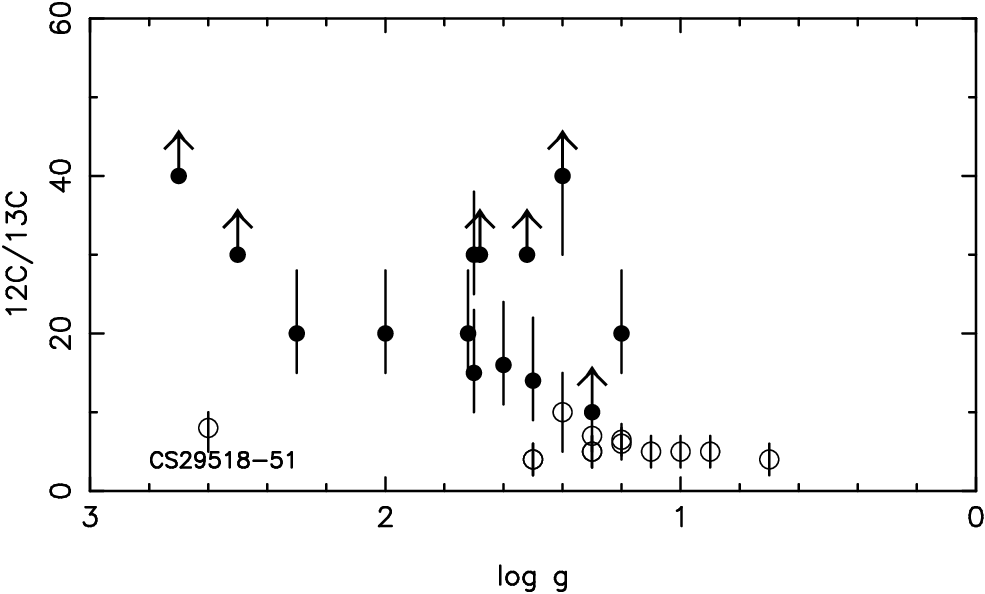} }
\caption{
${\rm ^{12}C/^{13}C}$ vs. Teff and log g for our sample of EMP giants.
Open and filled symbols denote ``mixed'' and ``unmixed'' stars,
respectively 
(see ``First Stars VI''). The lower ${\rm ^{12}C/^{13}C}$ ratios seen
at lower 
log g (i.e., higher luminosity) suggest extra mixing. Errors are
about 100K in 
\Teff~ and 0.2 dex in log g.
}
\label{c12c13}
\end {center}
\end {figure}

\begin {figure}[h]
\begin {center}
\resizebox  {8.0cm}{4.5cm}
{\includegraphics {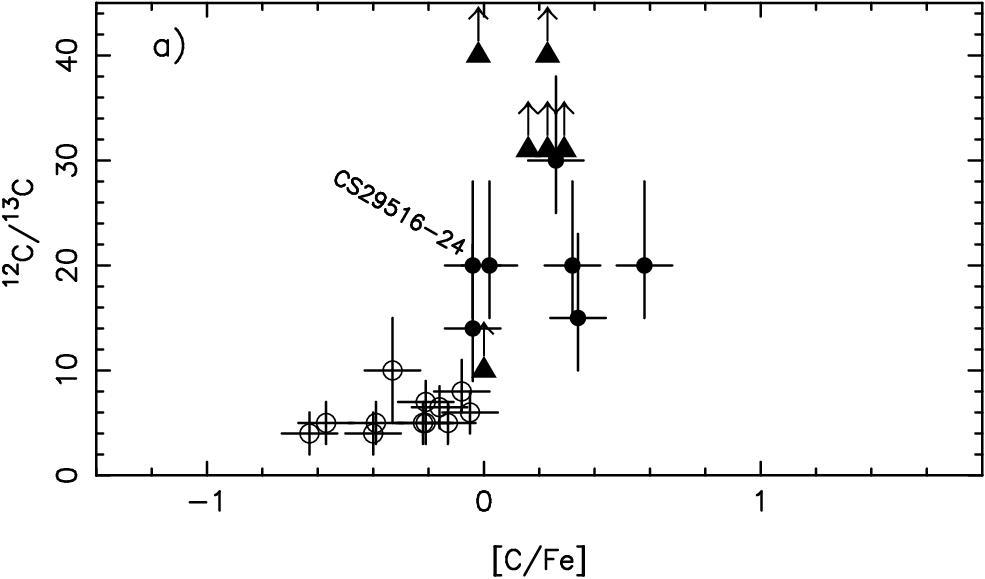} }
\resizebox  {8.0cm}{4.5cm}
{\includegraphics {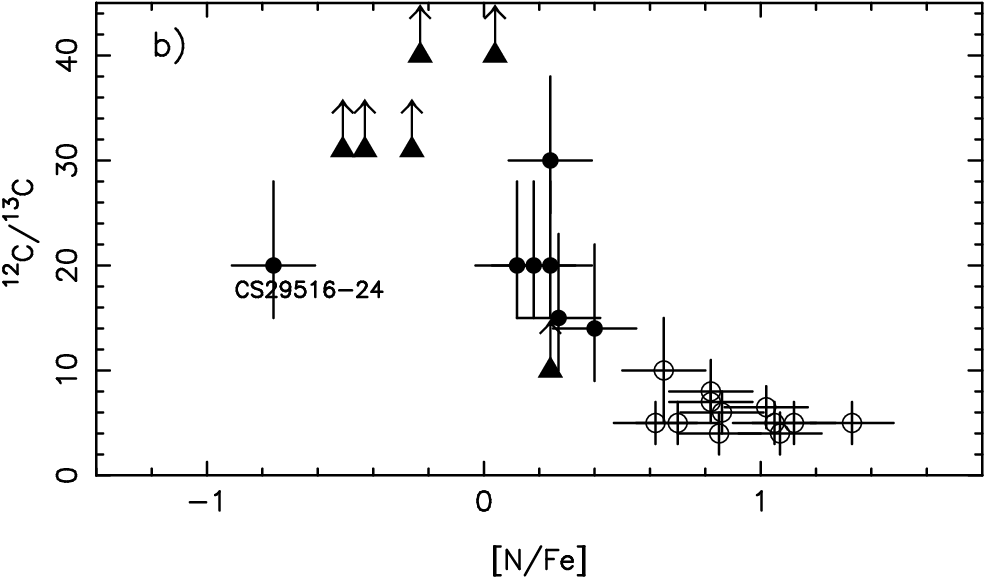} }
\resizebox  {8.0cm}{4.5cm} 
{\includegraphics {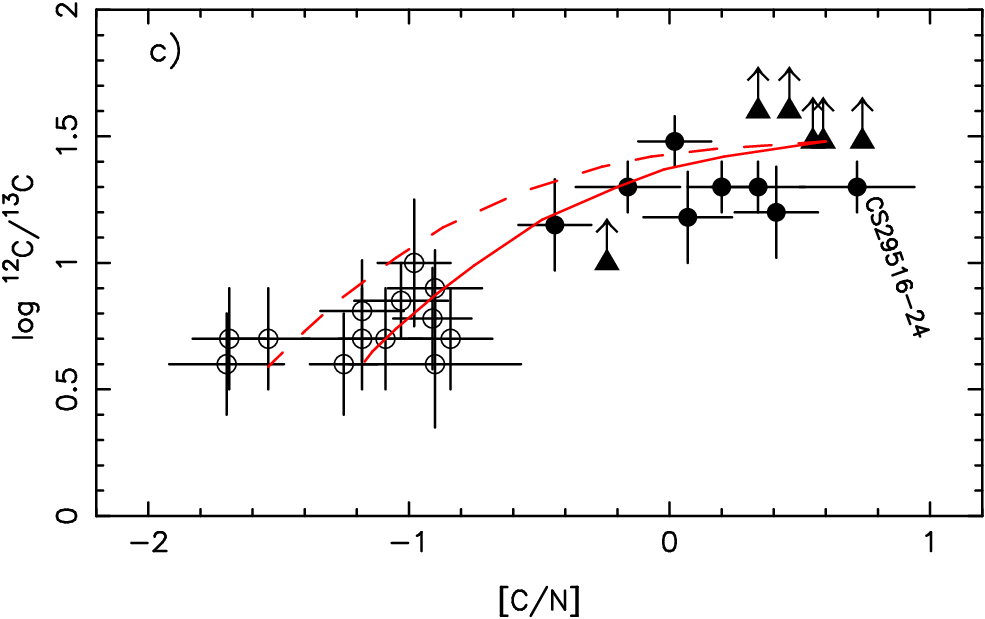} }
\caption {
$a$) ${\rm ^{12}C/^{13}C}$ vs.  [C/Fe], $b$) vs.  [N/Fe], and $c$)
${\rm log^{12}C/^{13}C}$ vs.  [C/N] for our sample of EMP giants.
Symbols are as in Fig.  \ref{c12c13}; triangles denote upper limits.
The correlation of ${\rm ^{12}C/^{13}C}$ with [C/Fe] (panel $a$) and
anticorrelation with [N/Fe] (panel $b$) are well defined; both could
be explained by extra mixing.  The low N abundance of CS~29516-024
presumably reflects the composition of the ISM from which it formed.
\hspace{1.0cm} The gap in [C/N] used to separate the mixed and unmixed
groups of stars is clearly visible in panel $c$, but a small gap also
appears within the mixed group, at ${\rm [C/N]\sim-1.4}$; the stars
with the lowest [C/N] values are also Na- and Al-rich (see
Sect.~\ref{Na-Al-rich} and Fig.~\ref{cn-alna}).  \hspace{1.0cm} The
curves show the computed evolution of ${\rm log^{12}C/^{13}C}$ as a
function of [C/N] in a stellar atmosphere progressively mixed with
material in which 80\% (full line) and 90\% (dashed) of the C has been
transformed into N by the CNO cycle.
}
\label{cnfe}
\end {center}
\end {figure}

In Fig.  \ref{c12c13} we show the ${\rm^{12}C/^{13}C}$ ratio vs.
\Teff\ and log g for our sample of EMP giants.  Clearly, all the stars
found to belong to the mixed subsample in ``First Stars VI'' (Spite et
al.  \cite{SCP05}) have a low ${\rm^{12}C/^{13}C }$ ratio.

In general, we found the unmixed giants to lie on the lower RGB, while
the mixed giants are on the HB or upper RGB. However, CS~29518-051
seems to belong to the low RGB (\Teff\ = 5200, log g = 2.6; see Fig.
\ref{lumin}), although we find it exhibits the characteristically high
[N/H] and low [C/H] and A(Li) of mixed giants (``First Stars VI'').
CS~29518-051 also has a very low ${\rm^{12}C/^{13}C}$ ratio,
confirming that its atmosphere has been mixed with CNO-processed
matter.  Since a large error in log g of this star is unlikely, it
would be interesting to monitor its radial velocity to verify whether
it may have been polluted by a binary companion.  As an alternative,
CS~29518-051 could be a horizontal-branch star (see Sect.
\ref{lumibump}).

\subsection {Correlations between ${\rm^{12}C/^{13}C}$ and the C and
N 
abundances} \label {s-c13cn}

Fig.  \ref{cnfe}a,b shows ${\rm^{12}C/^{13}C}$ vs.  [C/Fe] and [N/Fe]
for our stars (Table \ref{tabund}).  As seen, ${\rm^{12}C/^{13}C}$
increases steeply with [C/Fe] and decreases as [N/Fe] increases.
Moreover, we find ${\rm log(^{12}C/^{13}C) \approx 0.7}$ at the lowest
[C/N] ratios (i.e., most complete processing), close to the
equilibrium ratio for the CNO process.

The star CS~29516-024 has a rather low C abundance for an unmixed star
(Fig.  \ref{cnfe}a), but it also has the lowest [N/Fe] in our sample
(Fig.  \ref{cnfe}a), resulting in a normal [C/N] ratio (Fig.
\ref{cnfe}c).  Its relatively high ${\rm^{12}C/^{13}C = 20}$ (Fig.
\ref{cnfe}a-c) indicates that any mixing of the atmosphere with deeper
layers has not been very important, so the gas that formed this star
should have had the observed low N abundance, or perhaps even lower.

The effects of mixing with CNO processed material are illustrated in
Fig.  \ref{cnfe}c.  The curves show the relations between ${\rm
^{12}C/^{13}C}$ and [C/N] that result when a stellar atmosphere is
progressively mixed with material in which 80\% (full line) or 90\%
(dashed) of the initial C has been burned into N. Note that the three
leftmost stars in Fig.  \ref{cnfe}c (CS~22873--055, CS~22878--101, and
CS~22891--209) have lower [C/N] ratios than most of the other mixed
stars, which can be understood in terms of mixing with nearly fully
CNO processed material (dashed line).  CS~22952-015 should be added to
this list, as it exhibits the lowest C abundance and [C/N] ratio in
the sample; unfortunately, its C abundance is too low to yield a
usable ${\rm^{12}C/^{13}C}$ ratio, so it is not shown in the figures.
Most of these stars are found below (see Sect.~\ref{Na-Al-rich}) to
also show enhanced Na and Al abundances.

\subsection { ${\rm^{12}C/^{13}C}$ ratios vs. Li abundances }  \label
{s-c13li}

\begin {figure}[ht]
\begin {center}
\resizebox  {8.0cm}{4.5cm} 
{\includegraphics {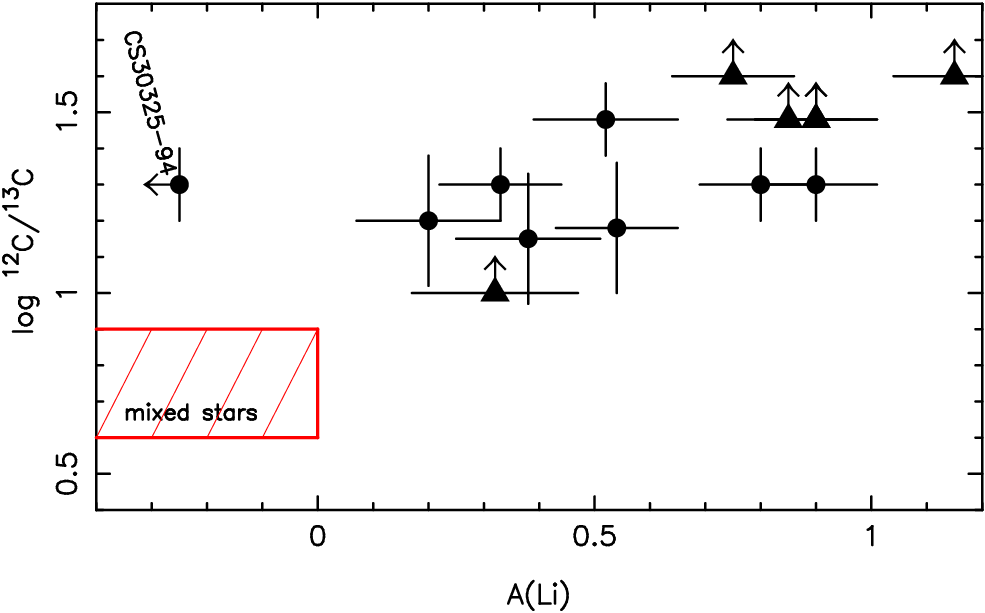} }
\caption {
${\rm log(^{12}C/^{13}C)}$ vs.  A(Li) (with A(Li)=${\rm
log~\varepsilon (Li) )}$; symbols as in Fig.\ref{cnfe}.  The dots are
compatible with a linear relation between ${\rm log^{12}C/^{13}C}$ and
A(Li) for the unmixed stars in the interval ${\rm 0.2 < A(Li) < 1.2}$.
CS~30325- 94 is an unmixed star with an unusually low
A(Li); see ``First Stars VI''.
The very low lithium abundance A(Li) cannot be measured 
in the mixed stars, but the upper limits all fall in the
hatched box in the diagram.
}
\label {c13li}
\end {center}
\end {figure}

Fig.  \ref{c13li} shows ${\rm log^{12}C/^{13}C}$ as a function of
A(Li) for our sample of unmixed giants. A(Li)=${\rm log~\varepsilon
(Li)}$, is the logarithm of the number of Li atoms per $10^{12}$ H
atoms.  In the mixed giants the Li line is too weak to be measured
(${\rm A(Li)<0.0}$), but ${\rm ^{12}C/^{13}C \approx 5.0 \pm 2}$; all
these stars fall in the hatched part of the diagram.  During the
evolution of a star, A(Li) decreases as the ${\rm ^{13}C}$ abundance
increases and ${\rm ^{12}C/^{13}C}$ gradually declines towards the CNO
equilibrium value.  Thus, the mixing process involves the whole
stellar envelope, bringing CNO processed material to the surface
layers while simultaneously depleting Li.

\section{Discussion}

\begin {figure}[t] 
\begin {center} 
\resizebox {7.1cm}{8.9cm}
{\includegraphics {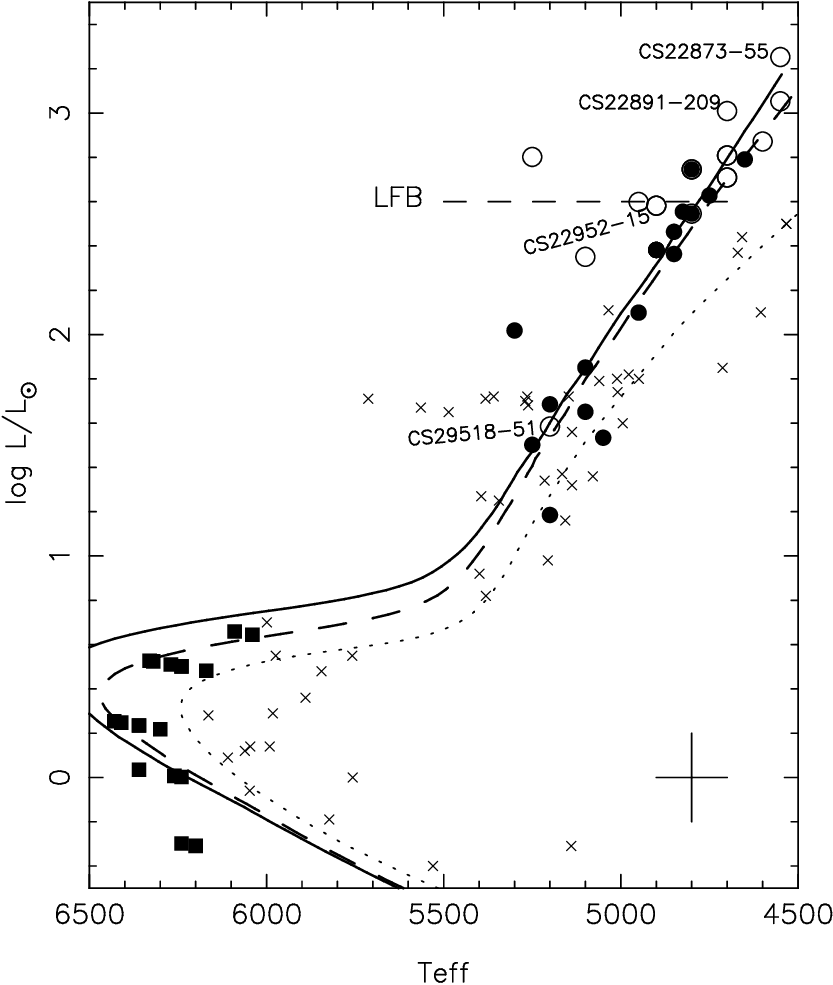} }
\caption {H-R diagram for our EMP dwarfs (from ``First Stars VII'';
filled 
squares), unmixed giants (filled circles), and mixed giants (open
circles). 
Crosses show the less metal-poor stars of Gratton et al. (2000). 14
Gyr 
isochrones by Kim et al. (2002) are shown for ${\rm[\alpha/Fe]=
+0.6}$, [Fe/H]= 
--3.7 (full line) and [Fe/H]= --2.7 (dashed line); a 12 Gyr isochrone
matching 
the Gratton et al. stars (${\rm[\alpha/Fe]= +0.3}$, [Fe/H]= --1.5) is
included 
for comparison (dotted line). The horizontal dashed line at ${\rm 
log~L/L_{\odot} \approx 2.6}$ indicates the luminosity function bump
of the 
giants (LFB) for ${\rm[Fe/H]\approx -3.2}$ (extrapolated from
Denissenkov \& 
VandenBerg, \cite{DV03}). The three labelled stars at the top of the
diagram are 
Na rich (Sect. \ref{AGB}). The mixed star CS~29518-051 could belong
to the HB 
instead of the lower RGB (Sect. \ref{analysis}).
}
\label{lumin}
\end {center}
\end {figure}

\subsection{The bump luminosity in metal-poor stars}  \label{lumibump}

Our EMP stars are too faint to have been observed by Hipparcos, so no
accurate direct distance and luminosity determinations are available
for them.  Meanwhile, we can estimate luminosities from their
atmospheric parameters via the classical formula: $${\rm
log~L/L_{\odot} = log~\mathscr{M/M}_{\odot} + 4~log~T_{eff} /
T_{eff,\odot} - log~g/g_{\odot}}$$
with ${\rm T_{eff,\odot}= 5780K}$, ${\rm log~g_{\odot}= 4.44}$ and
assuming all of the stars have a mass $\mathscr{M} = 0.85
\mathscr{M}_{\odot}$.

For the EMP giants, we take \Teff\ and log g from ``First Stars V'',
for the turnoff stars from Bonifacio et al.  (\cite {BMS06}, ``First
Stars VII'').  The resulting luminosities are listed in Table
\ref{tabund}.  With errors of 100K in \Teff~ and 0.2 dex in log g (see
``First Stars V''), we estimate that ${\rm \Delta (log~L/L_\odot)
\approx 0.2}$ dex.

Fig.  \ref{lumin} shows a comparison of our EMP giant and turnoff
stars with the stars observed by Gratton et al.  (\cite{GSC00}); as
expected, their moderately metal- poor stars (${\rm -2 <[Fe/H]< -1}$)
are on average cooler than our stars.  Fig.  \ref{lumin} also shows
isochrones by Kim et al.  (\cite{KDY02}) for 14 Gyr, ${\rm[\alpha/Fe]=
+0.6}$, [Fe/H]= --3.7 (full line) and [Fe/H]= --2.7 (dashed line).
For comparison, the dotted line shows an isochrone corresponding to
the Gratton et al.  stars (12 Gyr, ${\rm[\alpha/Fe]= +0.3}$, [Fe/H]=
--1.5; also from Kim et al.  \cite{KDY02}).  The isochrones fit the
data within the errors for both sets of RGB stars (the unmixed giant
CS~22966-057 seems to be overluminous for its \Teff\, hence it may be
a binary; see ``First Stars VI'').

As already noted by Denissenkov \& VandenBerg (\cite{DV03}), the
dwarfs of Gratton et al.  (\cite{GSC00}) appear to be shifted toward
lower temperature, although their mean metallicity is similar to that
of the giants.  A similar, but smaller, effect is seen for our EMP
turnoff stars, which tend to fall redward of the isochrone for [Fe/H]=
--2.7.  Following Denissenkov \& VandenBerg, the most likely cause is
a mismatch between the theoretical and the observed \Teff~ scales (see
on this point Mel\'endez et al.  \cite{MSV06}).

Gratton et al.  (\cite{GSC00}; their Fig.  7) found all of their
luminous giants (above ${\rm log~L/L_{\odot} \approx 1.8}$) to show
evidence of extra mixing between the surface and the H-burning layers,
i.e.:\\
$~~\bullet$ very low Li abundance (${\rm A(Li)} < 0.0$)\\ 
$~~\bullet$ high N abundance and low [C/N] ratio\\ 
$~~\bullet$ very low ${\rm^{12}C/^{13}C} $ ratio.\\ 

The luminosity ${\rm log~L/L_{\odot} \approx 1.8}$ corresponds to the
so-called ``bump'' in the luminosity function of the giants (LFB) in
the sample of Gratton et al.  (\cite{GSC00}).  At the LFB, the
hydrogen burning shell inside the star crosses the H-profile
discontinuity left behind by the base of the convective envelope
during the first dredge-up, and the $\mu$ gradient barrier disappears,
allowing the extra mixing to work.

The mixed stars in our sample present the same signatures (Li, C/N,
${\rm ^{12}C/^{13}C}$), and are identified in Fig.  \ref{lumin} by
open circles; extra mixing clearly sets in at a higher luminosity in
our more metal-poor EMP stars.

Theoretically, the luminosity of the LFB is expected to increase with
decreasing metallicity (Charbonnel \cite{Ch94}).  Denissenkov
\& VandenBerg (\cite{DV03}; their Fig.  1-2) give results for three
metallicities: log Z= --2.7, --3.0 and --3.3 (Z= 0.002, 0.001, 0.0005)
and $\mathscr{M} = 0.85 \mathscr{M}_{\odot}$.  The bump luminosity
appears to be an almost linear function of log Z; if we extrapolate it
to the mean metallicity of our stars (log Z= --4.52, Z= 0.00003), we
find the LFB to occur at ${\rm log~L/L_{\odot} \approx 2.6}$,
corresponding well to the onset of mixing in our EMP stars.  The link
between the extra-mixing phenomenon and the bump thus seems to be well
established.

Fig.  \ref{lumin} shows that, surprisingly, one mixed star,
CS~29518-051, is located well below the bump, very close to the
horizontal branch as defined by the HB stars of Gratton et al.
(\cite{GSC00}).  As the position of the HB depends only slightly on
metallicity, CS~29518-051 could be in fact an HB star which has
already gone through the RGB bump as well as the helium core flash.  Its
abundances, typical of mixed stars, could be due to extra-mixing processes
that occured during these phases.
(Note that the HB was accidentally misplaced in Fig.
9 of ``First Stars VI'').

\begin {figure}[t]
\begin {center}
\resizebox{8.0cm}{5cm} 
{\includegraphics {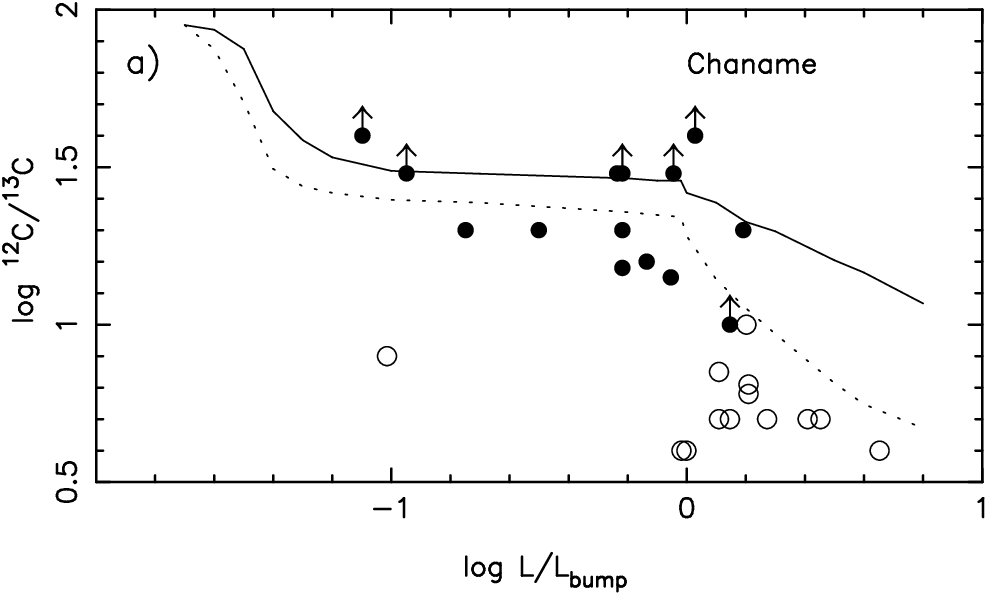} }
\resizebox{8.0cm}{5cm}
{\includegraphics {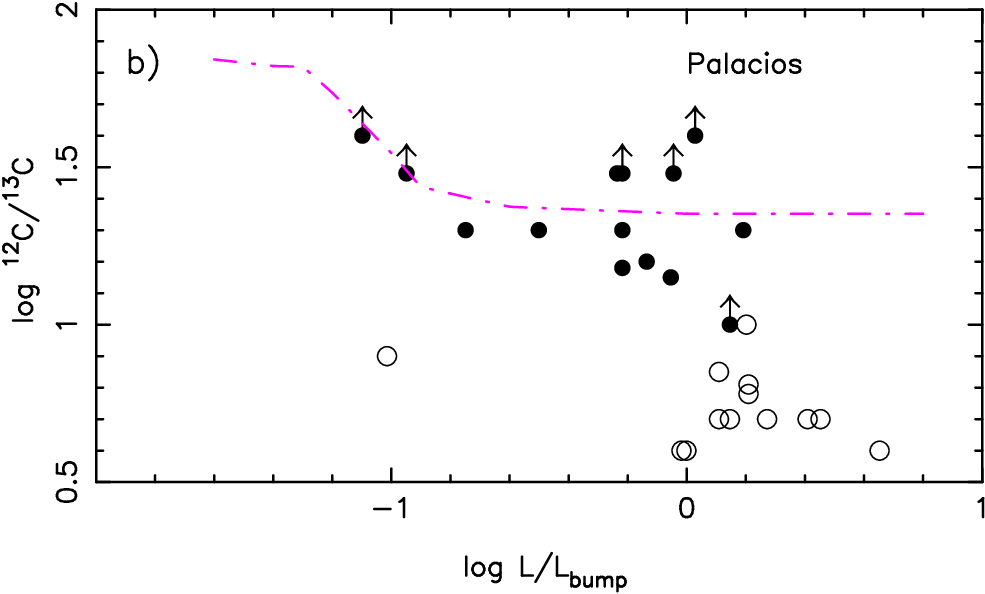} }
\resizebox{8.0cm}{5cm} 
{\includegraphics {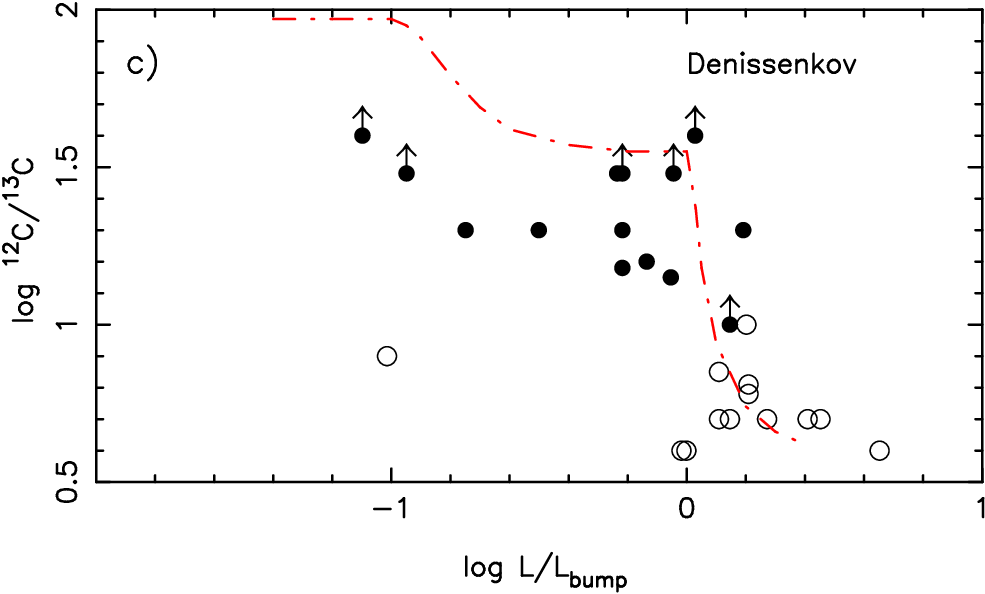} }
\caption {${\rm log^{12}C/^{13}C}$ vs. ${\rm log L/L_{bump}}$ in
mixed and 
unmixed giants (symbols as in Fig. \ref{c12c13}), compared to 
\hspace{0.5cm}
$a$) the predictions of Chanam\'e et al.  (\cite{CPT05}) for [Fe/H]=
--1.4 and ${\rm V_{TO}=}$ 40 \kms (full line) and 70 \kms (dotted
line).
\hspace{0.5cm}$b$) the predictions of Palacios et al. (\cite{PCT06})
which correspond to a value of  ${\rm V_{TO}}$ compatible to the 
observations (${\rm \approx 4 km s^{-1}}$).
The rapid drop in ${\rm ^{12}C/^{13}C}$ would
require very fast rotation speeds at the
turnoff, incompatible with the observations.
\hspace{0.8cm} $c$)  the predictions of Denissenkov \& Weiss 
(\cite{DW04}).  In this
case the rapid drop in ${\rm ^{12}C/^{13}C}$ at the RGB bump is
predicted.  However, this model is essentially a standard model on top of
which ad-hoc extra mixing is simulated.
}
\label{c13-chaname2}
\end {center}
\end {figure}

\subsection{Comparison with models of stellar structure}

\begin {figure}[ht]
\begin {center}
\resizebox  {8.0cm}{4.5cm}
{\includegraphics {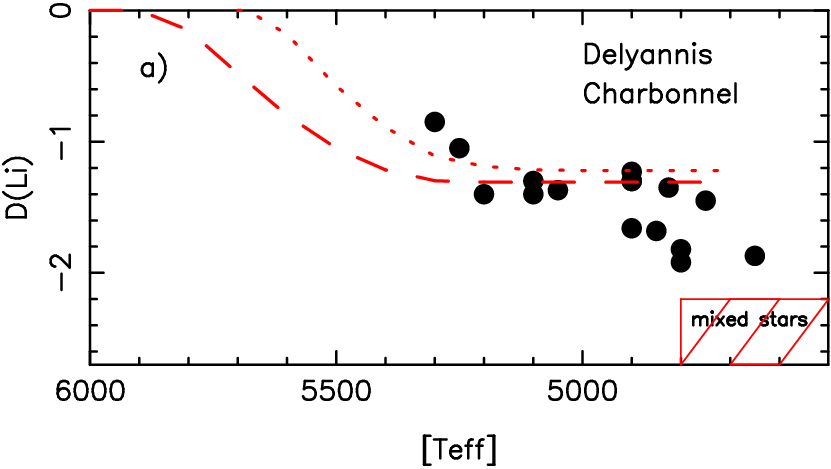} }
\resizebox  {8.0cm}{4.5cm} 
{\includegraphics {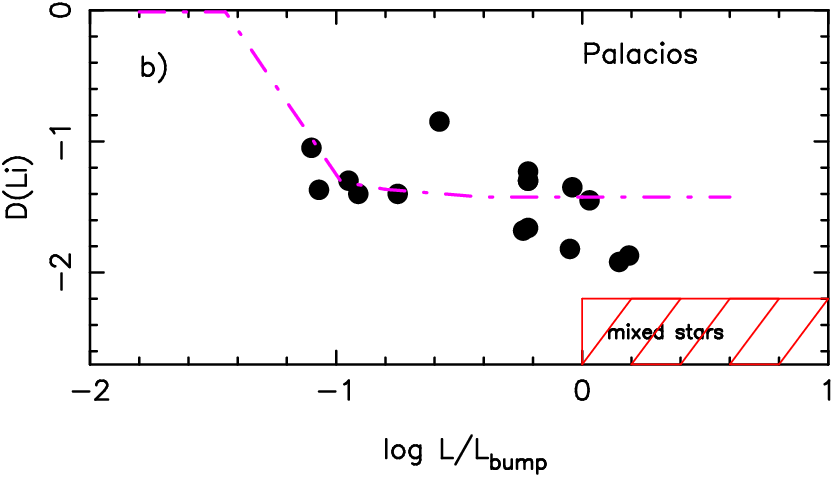} }
\resizebox  {8.0cm}{4.5cm} 
{\includegraphics {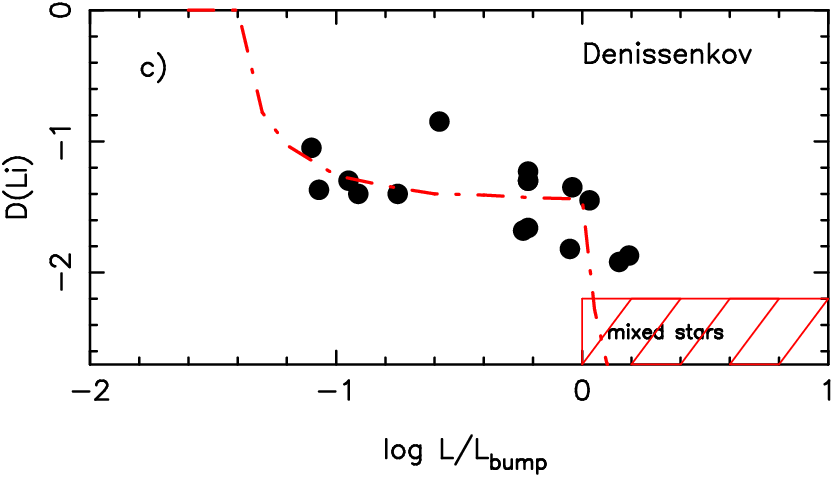} }
\caption {
Lithium depletions ($\rm{D(Li) = A(Li)-2.2}$) for our unmixed stars
(dots), compared with model predictions.  The hatched region contains
the mixed RGB stars, for which ${\rm D(Li) < -2.2}$. 
\hspace{0.5cm} $a$) Models by
Charbonnnel (dashed line) and Delyannis (dotted line), which do not
include extra mixing.  The observed Li depletion continues below
4900K, at variance with the predictions.  
\hspace{0.5cm}$b$) Self-consistent model (M5) of Palacios et al.
(\cite{PCT06}). 
\hspace{0.5cm}$c$) Model by Denissenkov \& Weiss (\cite{DW04}), which
include extra mixing explicitly (see Fig.6).
}
\label{dli}
\end {center}
\end {figure}

The ${\rm log ^{12}C/^{13}C}$ ratio is a powerful tool for further
tests of models of extra mixing in EMP stars, because it should depend
only on the internal mixing in the star and hardly at all on the
composition of the near- primordial ISM from which the stars formed.
By contrast, the atmospheric C and N abundances depend not only on
internal mixing, but also on the initial C and N abundances, which may
vary from star to star.

Fig.  \ref{c13-chaname2} shows ${\rm log ^{12}C/^{13}C}$ as a function
of ${\rm log ~L/L_{bump}}$ for our stars.  Clearly, as soon as a star
reaches the bump luminosity, ${\rm ^{12}C/^{13}C}$ drops very
abruptly.  Two classes of models have been developed in order to
account for these abundance changes:
\begin{itemize} 
\item The first class of models are the ``parametric diffusion
models."  Here, extra mixing is simulated by adding a diffusion term
to the equations that control the abundances of the different elements
at different levels inside the stars (Boothroyd \& Sackman
\cite{BS99}, Denissenkov \& VandenBerg \cite{DV03}).
\item In the second class of models, the mechanism responsible for the
extra mixing is associated with the transport of angular momentum in
the stellar interior and modelled explicitly.  Different mechanisms
have been proposed, such as magnetic fields, gravity waves, and
stellar rotation (see Charbonnel \cite{Ch95}, Chanam\'e et al.  \cite{CPT05}, 
Palacios et al. \cite{PCT06}).  
Stellar rotation is often considered as the most promising
non-standard mechanism to produce extra mixing, and its efficiency has
been explored by Chanam\'e et al.  (\cite{CPT05}).
\end{itemize}

Unfortunately, the decline of ${\rm ^{12}C/^{13}C}$ along the RGB has
only been modelled for metallicities above ${\rm[Fe/H] \approx -2}$, a
factor of 10 to 100 times higher than the stars we study here.  
In Fig.  \ref{c13-chaname2}a we have compared our data with the
predictions of Chanam\'e et al.  (\cite{CPT05}) for [Fe/H]= --1.4 and
turnoff rotational velocities of 40 and 70 \kms.  Neither of these
values can explain the rapid drop in ${\rm ^{12}C/^{13}C}$ for ${\rm
log ~L/L_{bump} > 0}$.\\
The best agreement is obtained with the higher rotational velocity, as
has also been found when less metal-poor giants are compared with
models. However, metal-poor field turnoff stars have rotational
velocities less than 10 \kms~ (see Lucatello \& Gratton 
\cite{LG03}).

The self-consistent models of Palacios et al.  \cite{PCT06} computed
for [Fe/H]=--1.6 do not represent better the observations (these
models correspond to rotational velocities at the turnoff compatible
to the observations i.e. ${\rm \approx 4.km s^{-1} }$).  In Fig.
\ref{c13-chaname2}b we compare our observations to the predictions of
the model M5 which gives the best fit, and none of the models predict
a rapid decrease of the ${\rm ^{12}C/^{13}C}$ ratio after the bump.

So we conclude that if the observed mixed stars really {\it are} on
the RGB, another extra-mixing process must be active.  Note that
Chanam\'e et al.  and Palacios et al. have assumed solid
rotation for the MS turnoff stars; if they have started their models
from differentially rotating MS stars, the extra mixing (following
Chanam\'e et al.  (\cite{CPT05}), would have been more vigorous for a
given rotation rate.

In Sect.  \ref {s-c13li} we found that a good correlation exists
between ${\rm ^{12}C/^{13}C}$ and A(Li).  It is interesting to check
whether the Li depletion in giants can also be explained by models.
Fig.  \ref{dli}a shows the Li depletion as a function of \Teff, taking
the initial ("plateau") Li abundance to be ${\rm A(Li)_{0} = 2.2}$
(thus ${\rm D(Li) = A(Li) - 2.2}$).  The unmixed stars, which have
undergone the first dredge-up, are shown individually; in the mixed
stars, we only have upper limits to A(Li), and these stars are located
inside or below the hatched box.\\
The observational results are compared with theoretical Li depletion
predictions following Charbonnel or Deliyannis (private
communications to Garc\'{\i}a P\'erez \& Primas \cite{GPP06}, see
references therein), computed for stellar masses of 0.85 and
$0.75M_{\odot}$, ${\rm [Fe/H] \approx -2}$, and without
extra-mixing.  The agreement is rather good for \Teff\ higher than
5000K, as the computed effect of the first dredge-up agrees with the
observations; but at lower temperatures, i.e. higher luminosities,
the Li abundance decreases more rapidly than predicted by the 
standard models.

In Fig.  \ref{dli}b the observed Li depletions are compared to the
predictions of the self consistent models of Palacios et al.
(\cite{PCT06}): these models are not able to represent the rapid
decrease of the lithium abundance after the bump; the computed effect of the 
rotation induced extra-mixing on the lithium abundance is negligible.

Finally in Fig. \ref{dli}c the observed Li depletions are compared to
the computations of Denissenkov \& Weiss (\cite{DW04}), which
explicitely include extra mixing.  The agreement is rather good, 
but the parameters of the extra mixing (depth and rate) are ad-hoc 
and its cause (rotation, gravity waves\ldots) is not defined.  Because
the same parameters can explain the lithium abundances 
and the ${\rm ^{12}C/^{13}C}$ ratios, we can expect that extra mixing 
will be able to predict the abundances in the extremely 
metal-poor RGB stars once its phsyical cause is better understood.

For a better understanding of the mixing processes in the EMP giants,
it would be important to extend the theoretical predictions of ${\rm
^{12}C/^{13}C}$ and D(Li) along the RGB to include stars with
${\rm[Fe/H] \leqslant -3.0}$.

\subsection{The mixed stars: RGB or AGB?}  \label{AGB}

So far, we have assumed that all our EMP giants lie on the RGB - the
unmixed stars on the lower RGB, the mixed stars on the upper RGB (see
``First Stars VI").  However, two facts advise caution -- the Na and
Al abundances and excess luminosity of some of the stars.  As Fig.
\ref{lumin} shows, some of the mixed stars are slightly more luminous
than the theoretical RGB branch, and these might be AGB stars.
Johnson (\cite{Joh02}) also found that four among 23 metal- poor stars
observed by her appeared to be AGB stars (as already noted by Bond
\cite{Bo80}).  In such stars, the core helium flash might perhaps
induce extra mixing.

Gratton et al.  (\cite{GSC00}) observed several metal-poor horizontal
branch (HB) stars, which have undergone the core helium flash.  These
stars
exhibited the same characteristics as the mixed RGB stars of their
sample (low C and Li and high N abundances, and low ${\rm
^{12}C/^{13}C}$ ratio).  Thus, the general abundance patterns of mixed
RGB and HB/AGB stars are rather similar.
 
\subsubsection{The Na and Al anomalies} \label{Na-Al-rich}

During the evolution of moderately metal-poor stars
(${\rm[Fe/H] > -2.0}$) in the field, from the main sequence to the top of the RGB,
Gratton et al.  (\cite{GSC00}) have not detected any variation of
[Na/Fe], as can be seen from inspection of their Figure 7.

In AGB stars with very low metallicity (${\rm Z=10^{-5}}$), the temperatures are
higher, and rotationally-induced deep-mixing episodes are expected to occur
(Meynet \& Maeder \cite{MM02}, Herwig \cite{Her05}), which can affect the atmospheric
Na and Al abundances.
Proton capture converts C and O into N, but also Ne into Na and Mg
into Al.  However, proton capture on Mg requires 
temperatures which are generally considered as too high for being 
reached in RGB stars. 
Following Weiss \& Charbonnel (\cite{WC04}), Al anomalies would
be a likely signature of AGB stars.

\begin {figure}[th]
\begin {center}
\resizebox  {8.0cm}{4.5cm}
{\includegraphics {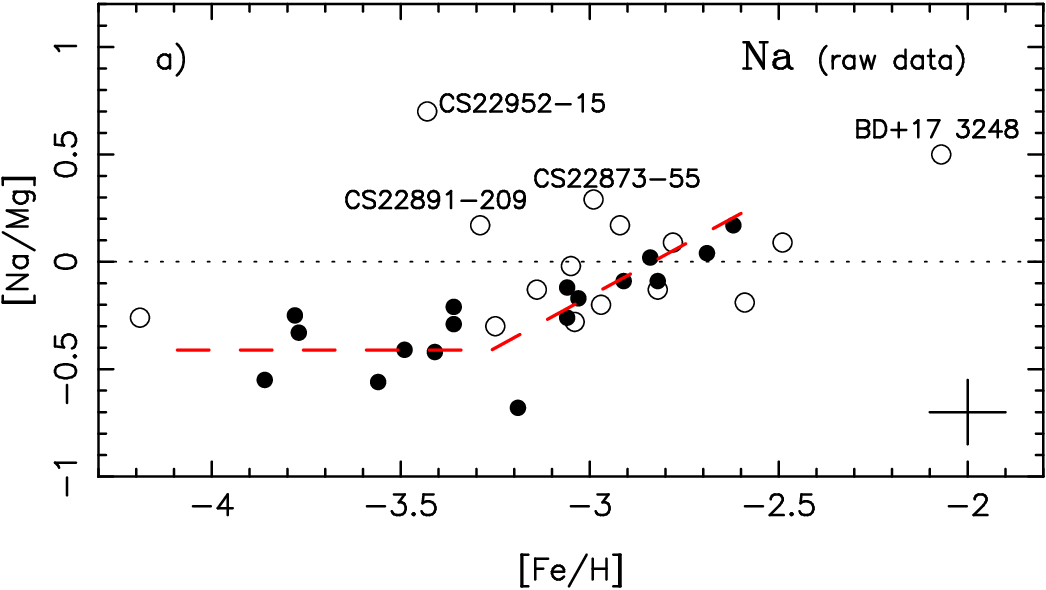} }
\resizebox  {8.0cm}{4.5cm} 
{\includegraphics {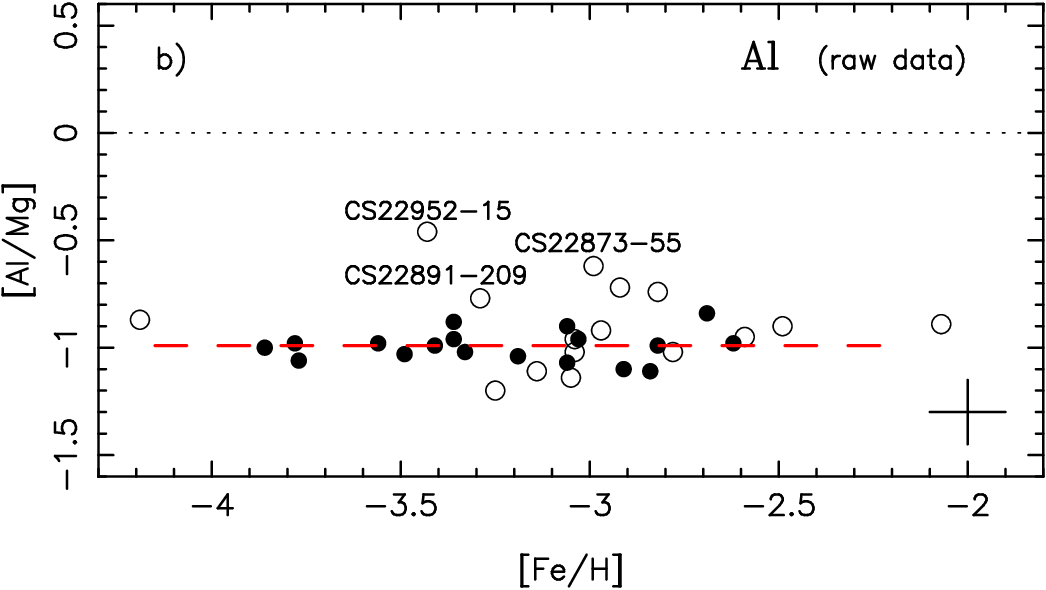} }
\caption {
[Na/Mg] and [Al/Mg] (without NLTE corrections, hereabove ```raw data")
vs.  [Fe/H]; symbols as in Fig.  \ref{c12c13}.  At low
metallicity (${\rm [Fe/H]<-2.6}$) all the Na-rich stars are mixed;
they are also Al-rich and located above the theoretical isochrone in
the H-R diagram (Fig.  \ref{lumin}), suggesting that they belong to
the AGB. The dashed lines represent the mean value of the
relations for the unmixed stars alone. The scatter in [Al/Mg] for
these stars alone is ${\rm \sigma_{[Al/Mg]} = 0.07}$ dex, consistent
with the measurement errors; thus, any cosmic scatter is below our
detection limit.
}
\label{alna}
\end {center}
\end {figure}

Fig.  \ref{alna} shows the observed [Na/Mg] and [Al/Mg] ratios vs.
[Fe/H].  No corrections for NLTE effects have been applied; while
Baum\"uller et al.  (\cite {BG97}, \cite {BBG98}) estimate that they
could reach --0.5 dex for Na and +0.6 dex for Al, Gratton et al.
(\cite {GCE99}) find a value of only --0.1 dex for Na.

\begin {figure}[th] \begin {center} 
\resizebox {8.0cm}{4.5cm}
{\includegraphics {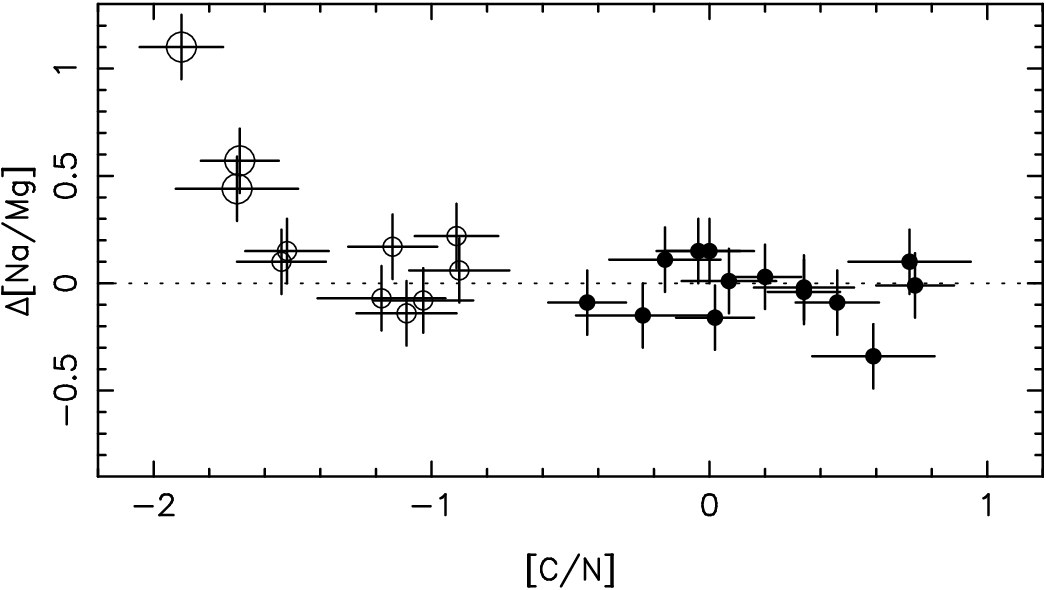} }
\resizebox {8.0cm}{4.5cm}
{\includegraphics {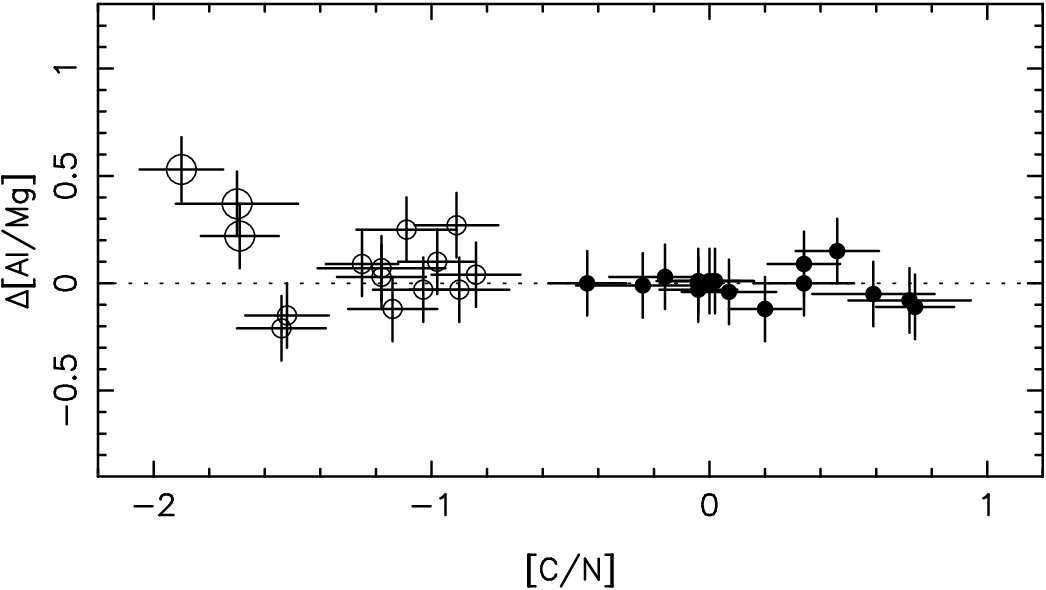} }
\caption { 
Na and Al excesses, ${\rm \Delta[Na/Mg]}$ and ${\rm \Delta [Al/Mg]}$,
relative to the relations shown by dashed lines in Fig.  \ref{alna},
vs.  [C/N], for ${\rm[Fe/H]<-2.6}$; symbols as in Fig.  \ref{alna}.  
The large open circles
denote the three Na- and Al-rich stars from Fig.  \ref{alna}; these
stars also have the lowest C/N ratios.
}
\label{cn-alna}
\end {center}
\end {figure}

Defining mean relations for the unmixed stars in Figs.  \ref{alna}a,b
(dashed lines), we can compute individual Na and Al excesses ${\rm
\Delta[Na/Mg]}$ and ${\rm \Delta [Al/Mg]}$ for all our stars.  These
are shown in Fig.  \ref{cn-alna} as functions of [C/N]. 
Because [Na/Mg] increases with [Fe/H] (Fig.  \ref{alna}a) and because
there are no unmixed giants above ${\rm [Fe/H] > -2.6}$ in our sample,
we cannot compute mean values of [Na/Mg] and corresponding Na and Al
excesses for ${\rm [Fe/H] > - 2.6}$.  Thus, the most metal-rich stars
in our sample are not included in Fig.  \ref{cn-alna}.

The mixed stars have slightly higher Na and Al excesses than the
unmixed stars.  Three stars with ${\rm [Fe/H] < -2.6}$ are both Na-
and Al-rich, and all three are mixed stars (Fig.  \ref{alna} \&
\ref{cn-alna}): CS~22873-55, CS~22891-209, and CS~22952-15.
These three stars all have low [C/N] ratios (${\rm
[C/N]<-1.6}$); CS~22952-015 has the lowest [C/N] ratio of them all and
is also Mg-poor, as expected if Mg-Al cycling has played an important
role.

In summary, the evidence suggesting that the Na- and Al-rich stars are
in fact AGB stars are:
\begin{itemize}
\item The observed Na and especially Al enrichment requires mixing
with very deep high-temperature layers, which is not predicted to
occur in RGB stars;
\item The Na-Al-rich stars are located above the theoretical RGB
isochrone (Fig.~\ref{lumin}).
\end{itemize}

Note that the most metal-rich star of our sample in Fig. \ref{alna} is
a mixed star: ${\rm BD+17^{o}3248}$ ([Fe/H]= --2.07).  Following the
measurements of Johnson (\cite{Joh02}), the relations [Na/Mg] and
[Al/Mg] vs.  [Fe/H] are flat for ${\rm -2.5 < [Fe/H] < -2.0}$ and thus ${\rm
BD+17^{o}3248}$ could also be Na-rich.  Its aluminum abundance
seems to be normal.  At $\rm{ log~L/L_{\odot} = 2.80}$ and \Teff= 5250
K, this star is far above the isochrones in Fig.~\ref{lumin}, and could
thus also be an AGB star, as already claimed by Bond (\cite{Bo80}) and
Johnson (\cite{Joh02}).  However, note that its [C/N] ratio, while low
([C/N]= --0.98), is not extreme.

Unlike most AGB stars, the three Na-rich EMP stars in our sample are
not enriched in $s$- elements such as Sr or Ba (Fran\c cois et al.,
\cite{FDH06}, ``First Stars VIII'').  According to Cowan et al.
(\cite{CSB02}), ${\rm BD+17^{o}3248}$ is very rich in neutron-capture
elements, but these seem to have been produced by the $r$-process,
probably during an earlier supernova explosion, not in the star
itself.  The $s$-element production in AGB stars occurs quite late in
the AGB phase, after the start of thermal pulses; thus, we would not
expect to see an enhancement of these elements if the observed stars
are in the early AGB phase.

However, alternative interpretations of the Al and Na excesses are
possible, in particular when one takes into account
the fact that the temperatures 
needed in particular to transform Mg into Al are encountered only in 
AGB stars; 
e.g.:
\begin{itemize}
\item mass transfer from a former, more massive AGB binary companion,
which has now evolved into a white dwarf.  It would be very
interesting to monitor the radial velocities of these stars to check
this hypothesis.
\item inhomogeneous enrichment of the ISM by a previous generation of
AGB stars.
\end{itemize}

Finally, it is interesting to note that the mean value of [Al/Mg] over
all stars (Fig.  \ref{alna}b) is ${\rm <[Al/Mg]>= -0.95 \pm 0.15}$
(s.d.), but the scatter drops by a factor two if {\em only the unmixed
stars} are considered: ${\rm <[Al/Mg]>= -0.99 \pm 0.07}$ (no NLTE
corrections).\\
A similar trend is seen for Na, but ${\rm <[Na/Mg]>}$ only becomes
constant below [Fe/H] = --3.2, and the scatter remains higher than for
Al (Fig.  \ref{alna}a).

\subsubsection{Are signatures of the ON cycle visible in the mixed stars ?}

Because we seem to observe products of the Ne-Na and Mg-Al cycles in
some stars, we might expect products of the ON cycle to be directly
observable as well, since the ON cycle occurs at about the same 
temperature as the NeNa chain (e.g. Weiss \& Charbonnel \cite{WC04}).

\begin {figure}[ht]
\begin {center}
\resizebox  {8.0cm}{4.5cm} 
{\includegraphics {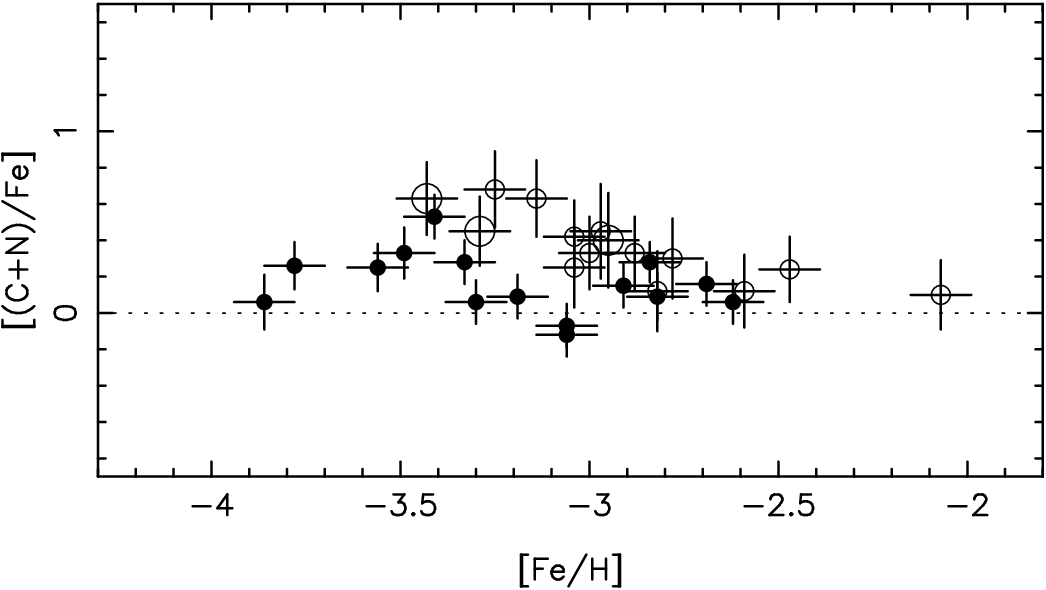} }
\caption {
[(C+N)/Fe] vs. [Fe/H]; symbols as in Fig. \ref{alna}. The three large
open 
circles indicate the three Na- and Al-rich stars. The C+N excess of
the mixed 
stars could be due to the ON cycle.
}
\label{c+n}
\end {center}
\end {figure}

\begin {figure}[ht]
\begin {center}
\resizebox  {6.0cm}{6.0cm}
{\includegraphics {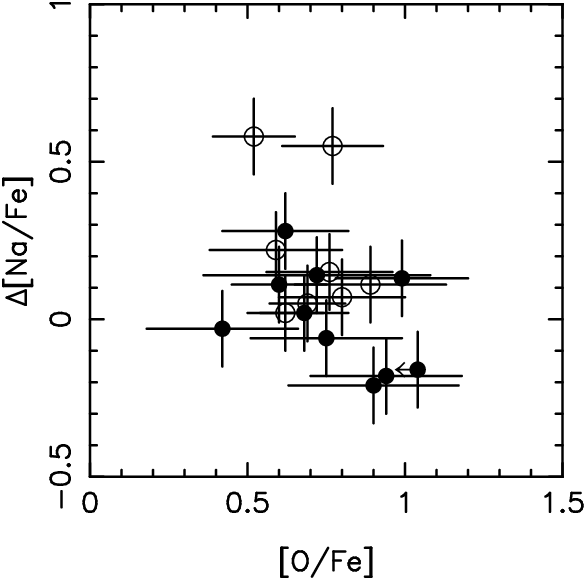} }
\caption{
$\Delta$~[Na/Fe] vs. [O/Fe]; symbols as in Fig. \ref{cnfe}.
$\Delta$~[Na/Fe] is 
the Na excess relative to ${\rm <[Na/Fe]>}$ at the [Fe/H] of the
star. There is 
no obvious correlation between Na excess and [O/Fe].
}
\label{dna-o}
\end {center}
\end {figure}

Our earlier plot of [O/Fe] vs.  [Fe/H] (``First Stars VI'', Fig.  10)
showed no significant difference between mixed and unmixed stars.
Thus, we concluded that mixing with ON-cycled material is not
efficient.  However, the absolute abundance (i.e., number of atoms per
$10^{12}$ H atoms) is much larger for O than for N; thus, a decrease
in O due to ON cycling could exist, but might not be detected in
[O/Fe] or [O/Mg] due to their rather large observational errors.
Indeed, a plot of [(C+N)/Fe] vs.  [Fe/H] (Fig.  \ref{c+n}) seems to
show that, in the mean, the mixed stars have a C+N excess which could
be due to N produced by the ON cycle.

In the red giants of M5, an increase of C+N with decreasing C was also
observed by Cohen et al.  (\cite {CBS02}), and interpreted as the
result of an admixture of ON-processed material to the surface layers.
However, prudence is advisable in interpreting the [(C+N)/Fe] excess
in our mixed stars.  Because the N abundances of the mixed and unmixed
stars differ by almost a factor of ten, the total number of C+N atoms
is close to the number of C atoms alone in the unmixed stars, and to
the number of N atoms in the mixed stars.  Therefore, slight
systematic errors in the C or N abundance determinations (NLTE,
3D\ldots) could produce the observed effect.

For example, we recall that N abundances derived from the NH band
needed a correction of 0.4~dex to reconcile them with those from the
CN band (``First Stars VI'').  Had we instead adopted a correction of,
say, 0.8~dex, the mean value of [(C+N)/Fe] would be the same in the
mixed and unmixed stars, and there would be no indication of any ON
cycling.

We can also compare the scatter in the relations of [(C+N)/Fe] vs.
[Fe/H] and [(C+N+O)/Fe] vs.  [Fe/H].  If ON-processed material has
been mixed to the surface in some stars, the first relation should
exhibit larger scatter than the second one.  However, first, the
number of stars with [(C+N+O)/Fe] is smaller (O could be measured in
only 23 stars, two of them being peculiar C-rich stars).  Secondly,
because O is more abundant than C and N, the dispersion in
[(C+N+O)/Fe] is mainly due to the uncertainty of the O abundance,
which is large because the [O I] line is very weak.  Definite
conclusions on the presence of ON cycled matter cannot be drawn from
the scatter in these two relations.

Finally, if the chemical composition of the atmosphere of some giants
has been affected by the ON cycle, we could expect a relation between
[O/Fe] and Na excess.  This anticorrelation is observed both on the
RGB and in turnoff and subgiant stars of some globular clusters (e.g.,
Gratton et al.  \cite{GSC04}).  The best explanation of the anomalies
seen in globular clusters is that proton capture converted C and O
into N and Ne into Na, either
\begin{itemize}
\item in the H burning shell of the star itself, followed by
convection (not expected in giants, and impossible in turnoff or
subgiant stars);
\item or in a previous stellar generation which enriched the ISM 
inhomogeneously;
\item or in a former AGB binary companion, which polluted the
atmosphere of the 
observed star through stellar winds.
\end{itemize}

Fig.  \ref{dna-o} shows $\Delta$[Na/Fe] as a function of [O/Fe], where
$\Delta$[Na/Fe] is the Na excess relative to the average [Na/Fe] for
the [Fe/H] of the star (analogous to Fig.  \ref{cn-alna}, but for
[Na/Fe] instead of [Na/Mg]).  No significant anticorrelation between
$\Delta$[Na/Fe] and [O/Fe] is seen within the measurement errors.  The
two Na-rich stars in the figure, CS~22873-055 and CS~22891-209 (see
also Fig.  \ref{alna}), are probably AGB stars.  We have been unable
to measure O in the third Na-rich star, CS~22952-015, because its
radial velocity is small and the stellar [O I] line is obliterated by
a saturated sky emission line.

In summary, we cannot prove that an O-Na anticorrelation exists in the
mixed EMP giants, and thus have no clear evidence from this relation
that the ON cycle has been active.  However, because the Ne-Na and
Mg-Al cycles operate at the same or even a higher temperatures than the 
ON cycle (Weiss
\& Charbonnel \cite{WC04}), and we do find Na and Al enhancements in
a few stars, it is likely that the ON cycle has polluted the atmosphere
of at least these few mixed stars.

\section {Conclusions}

We have analysed new ${\rm ^{12}C/^{13}C }$ isotopic ratios for a
sample of 35 EMP red giants, including 22 stars with ${\rm[Fe/H]
\leqslant -3.0}$, and find the following new results:

$\bullet$ Clear (anti)correlations exist between ${\rm ^{12}C/^{13}C}$
and [N/Fe] and [C/Fe], as expected if the C and N abundances are
modified by admixtures of CNO-processed material.  However, there seem
to be large variations in the C and N abundance of the near-primordial
ISM from which these stars formed; many EMP stars are C-rich and some
are N-rich, so the C/N ratio is not a clean indicator of internal
mixing.  The ${\rm ^{12}C/^{13}C}$ ratio is a much better diagnostic,
because ${\rm ^{12}C/^{13}C}$ should be high in primordial matter
($>70$) and its determination is insensitive to the choice of
atmospheric parameters for the stars.

$\bullet$ The depletion of Li is the first observational signature of
mixing, because it occurs in shallower layers than the CNO cycle;
therefore, it is a useful complement to the C and N results which were
used in ``First stars VI'' to separate our sample into ``mixed'' and
``unmixed'' stars.  In the lower RGB stars, we find a narrow
correlation between ${\rm ^{12}C/^{13}C}$ and A(Li); in the more
evolved (mixed) stars, ${\rm log ^{12}C/^{13}C \approx 0.7}$ and Li 
(${\rm A(Li)<-0.1}$) is
not detected, and is likely to have been destroyed.

$\bullet$ Computing luminosities from \Teff\ and log g allows us to
place the stars in the H-R diagram.  For the RGB stars, we find a good
fit between the observations and 14 Gyr isochrones by Kim et al.
(\cite{KDY02}) for the appropriate chemical composition. Interestingly
a few mixed stars fall somewhat above the isochrone, suggesting an AGB
phase.

$\bullet$ The main signatures of deep mixing are a low ${\rm
^{12}C/^{13}C}$ ratio, a low Li abundance, and a high N abundance
(${\rm ^{12}C/^{13}C<10}$, ${\rm log A(Li) < 0.0}$, and ${\rm [N/Fe] >
0.5}$).  These appear above a luminosity of ${\rm log~L/L_{\odot}=
2.6}$.  This is higher than the luminosity at which extra mixing was
found in less metal-poor stars by Gratton et al.  (\cite{GSC00}):
${\rm log~L/L_{\odot}= 2.0}$.  However in both samples this limit of
deep mixing corresponds to the luminosity of the bump.

$\bullet$ At variance with the suggestion of Gratton et al.
(\cite{GSC00}) we did not find that the extra mixing decreases in
stars below ${\rm [Fe/H]\approx-1.5}$.

$\bullet$ Assuming that our mixed EMP stars are evolving along the
RGB, we find that extra mixing sets in when a star crosses the bump in
the luminosity function.  As soon as this happens, the ${\rm
^{12}C/^{13}C}$ ratio decreases very rapidly, both in EMP and in
moderately metal-poor stars.  Standard models that include only
convection can explain neither this observation nor the depletion of
Li at low temperature or high luminosity.\\
--Our results about ${\rm ^{12}C/^{13}C}$ are consistent with the
rotating models by Chanam\'e et al.  (\cite{CPT05}) for
globular-cluster stars, but only for very fast turnoff rotation rates
(${\rm V_{TO}>}$ 70 \kms).  Because metal-poor field turnoff stars
have ${\rm vsini<}$ 10 \kms, rotation is probably not the only cause
of extra mixing in RGB stars, but it would be important to check this
directly with models for EMP stars (${\rm [Fe/H] < -3}$).\\
--Our lithium measurements agree neither with the models of Delyannis 
and Charbonnel (as reported by Garci\'a P\'erez \& Primas, 
\cite{GPP06}) nor with the models of Palacios et al. (\cite{PCT06}).

$\bullet$ As an alternative interpretation, at least some of our mixed
stars could be early AGB stars, with the core helium flash driving
additional mixing processes.  Overall, there are no striking abundance
differences between mixed RGB stars and AGB stars, but the Na and Al
abundances offer useful clues.  For the first time in metal-poor field
giants, we find several of our mixed stars to be Na- and Al-rich,
suggesting that a deep-mixing episode has taken place.  This is
sometimes expected to occur in AGB stars (Herwig \cite{Her05}), but
Al enhancement seems to be impossible in RGB stars, even with extra mixing (e.g.
Weiss \& Charbonnel \cite{WC04}).  The Na- and Al-rich giants
generally lie above the theoretical RGB, suggesting that they may
indeed be AGB stars.  However, it remains possible that the Na and Al
(and N) enrichment could be due to inhomogeneous enrichment of the ISM
by a previous generation of stars, or to pollution by a former AGB
binary companion.

$\bullet$ The scatter of [Al/Mg] around its mean value for the unmixed
stars alone is extremely low: $\sigma =0.07$ dex, corresponding to the
measurement errors alone (see ``First Stars V").  The mean value of
[Al/Mg] in the pristine ISM appears to be ${\rm [Al/Mg]\approx-0.39}$
(applying an NLTE correction of +0.6 dex), with negligible cosmic
scatter.

$\bullet$ Because the Ne-Na and Mg-Al cycles seem to have influenced
some of our stars, we would expect that signs of the ON cycle should
also be visible in at least these stars.  It does indeed seem that, on
average, [(C+N)/Fe] is larger in the mixed stars than in the unmixed
stars.  However, this could perhaps be due to slight systematic errors
in the C or N abundances (e.g., from NLTE or 3D effects).

\vspace{0.3cm}
Understanding the abundances of light elements such as C, N, O, Na,
and Al and their scatter in the pristine ISM remains an important goal
for the future.  In order to make progress from abundance analyses of
EMP giants, it is crucial to understand what mixing processes may have
affected their atmospheres.  Dwarf and turnoff stars are free from
such mixing, but measuring N abundances in EMP dwarfs is practically
impossible, because even the near-UV band of NH is too weak unless the
stars are actually N rich.  Thus, EMP giants will remain a key source
of data in the foreseeable future, and the lines of analysis presented
will need further refinement.

For this to become possible, more accurate laboratory data for the NH
and CN bands, in particular, are badly needed.  NLTE 3D computations
are the next crucial step.  Precise trigonometric parallaxes for EMP
giants (from SIM or GAIA) will be essential in order to distinguish
RGB and AGB stars.  Finally, the radial velocities of the mixed giants
should be monitored to assess the frequency of binaries among them,
and thus whether the observed abundance anomalies could originate in a
companion star that has gone through the AGB stage in the past.

\begin {acknowledgements}
We thank the ESO staff for assistance during all the runs of our Large
Programme.  R.C., P.F., V.H., B.P., F.S. \& M.S. thank the PNPS and
the PNG for their support.  PB and PM acknowledge support from the
MIUR/PRIN 2004025729\_002 and PB from EU contract MEXT-CT-2004-014265
(CIFIST).  T.C.B. acknowledges partial funding for this work from
grants AST 00-98508, AST 00-98549, and AST 04-06784 as well as from
grant PHY 02-16783: Physics Frontiers Center/Joint Institute for
Nuclear Astrophysics (JINA), all from the U.S. National Science
Foundation.  BN and JA thank the Carlsberg Foundation and the Swedish
and Danish Natural Science Research Councils for partial financial
support of this work.
\end {acknowledgements}

\end{document}